\documentclass[11pt]{article}
 \usepackage{geometry}                
\usepackage{graphicx}

\usepackage{amsfonts, amsbsy, amssymb, amsmath, float, subfigure, color}
\usepackage{subfigure}
\usepackage{epstopdf}
\usepackage{multirow}
\usepackage{pdflscape}
\usepackage[utf8]{inputenc}
\usepackage{subfigure} 
\usepackage{todonotes}
 \usepackage[footnotesize]{caption}

\begin{document}
\title{\textbf{Lagrangian study of the final warming in the southern stratosphere during 2002: Part II. 3D structure.}}

\author{J. Curbelo       \and
        C. R. Mechoso \and
        A. M. Mancho \and
        S. Wiggins} 

\date{}

\maketitle

\begin{abstract}

This two-part paper {aims to provide} a Lagrangian perspective of the final southern warming in spring of 2002, during which the stratospheric polar 
vortex (SPV) experienced a unique splitting. We approach the subject from a dynamical systems viewpoint and search for Lagrangian coherent structures using a 
Lagrangian descriptor that is applied to reanalysis data. Part I presents our methodology and focuses by means of a kinematic model, on the understanding of  
fundamental processes for filamentation and ultimately for vortex splitting on an isentropic surface in the middle stratosphere. The present Part II discusses 
the three dimensional (3D) evolution of the flow during the {selected} event. For this, we apply concepts developed in Part I concerning a definition of 
the vortex boundary that helps in the selection of trajectories to illuminate the  evolving flow structures, and a criterion that allows to {justify why at 
an isentropic level a pinched vortex will split in later times.} Lagrangian structures identified include surfaces that are several kilometers deep, and which 
a particle trajectory analysis confirms as barriers to the flow.  The role of Lagrangian structures in determining the fate of particles during the SPV 
splitting is discussed.

.

\end{abstract}

\begin{figure*}
\begin{center}
\begin{tabular}{cccc}
&22 Septiembre &  24 September 2002 & 26 September 2002\\
  \rotatebox{90}{\hspace{2cm} 40 km}& 
 \includegraphics[scale=0.3]{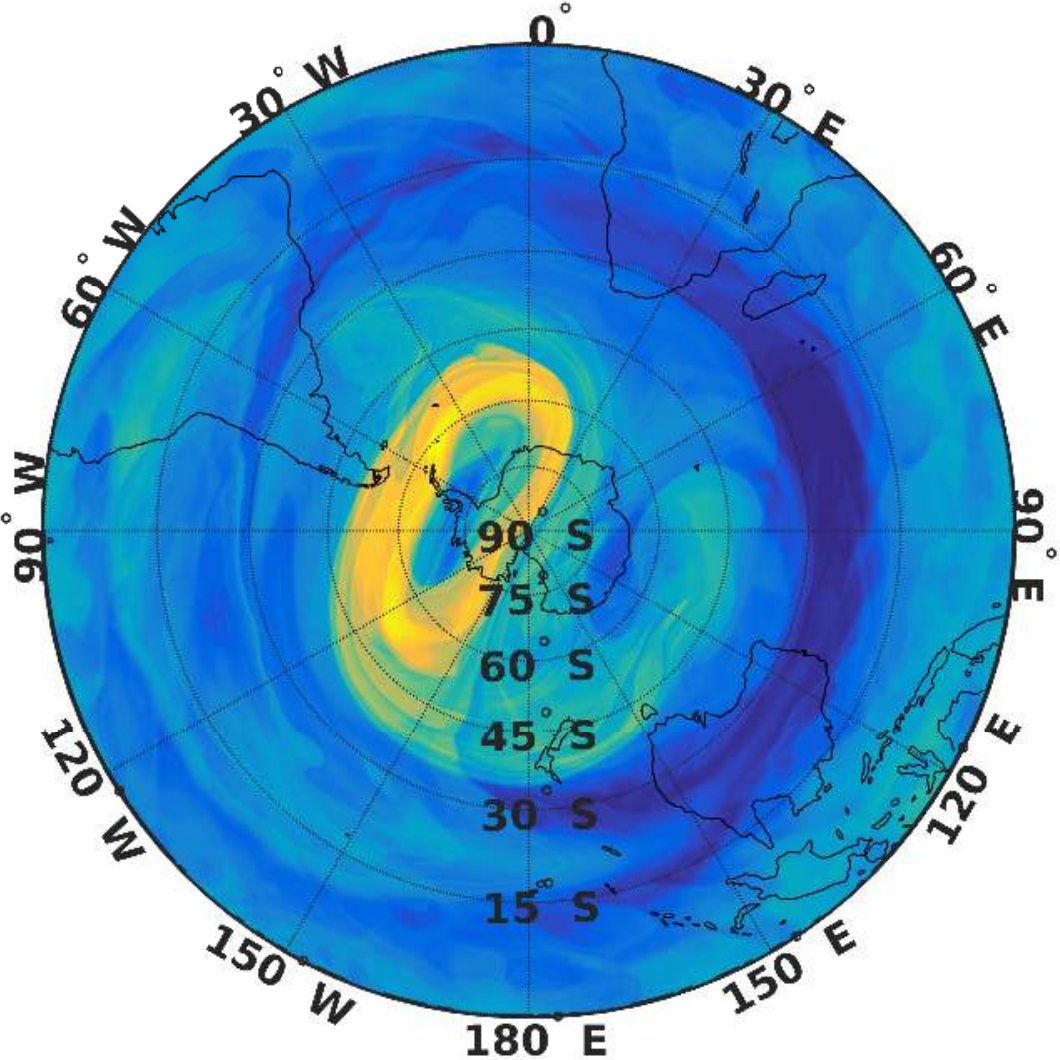}&  
\includegraphics[scale=0.3]{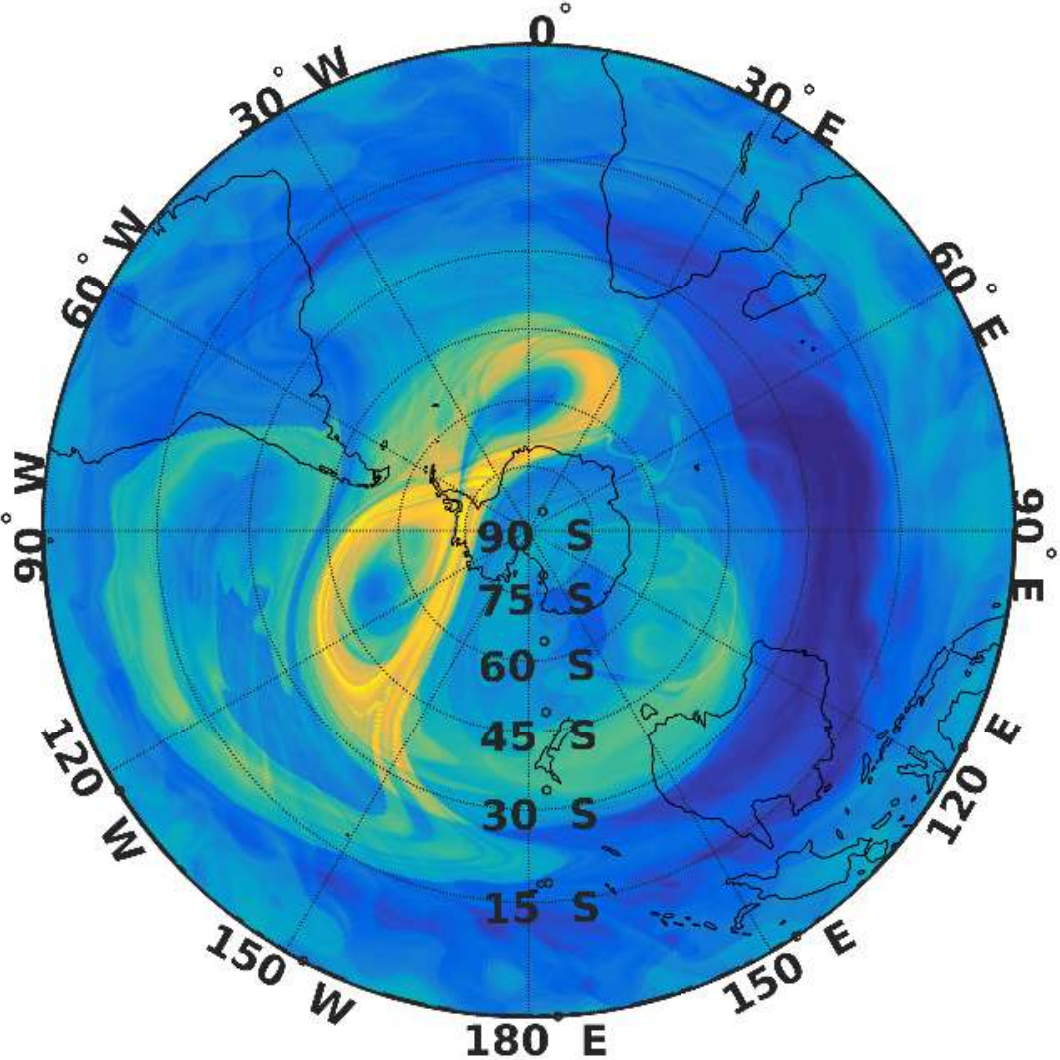}&\includegraphics[scale=0.3]{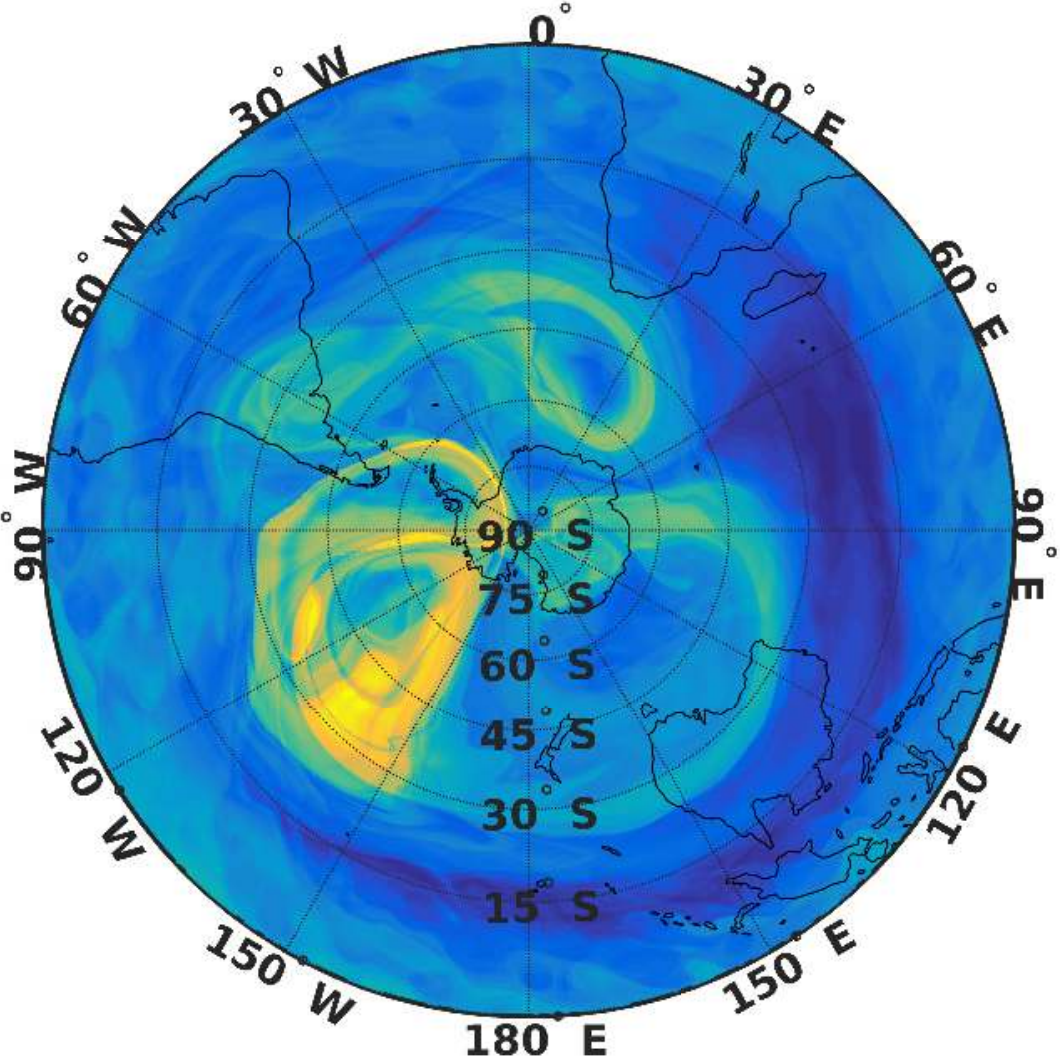}\\
   \rotatebox{90}{\hspace{2cm} 30 km}& 
\includegraphics[scale=0.3]{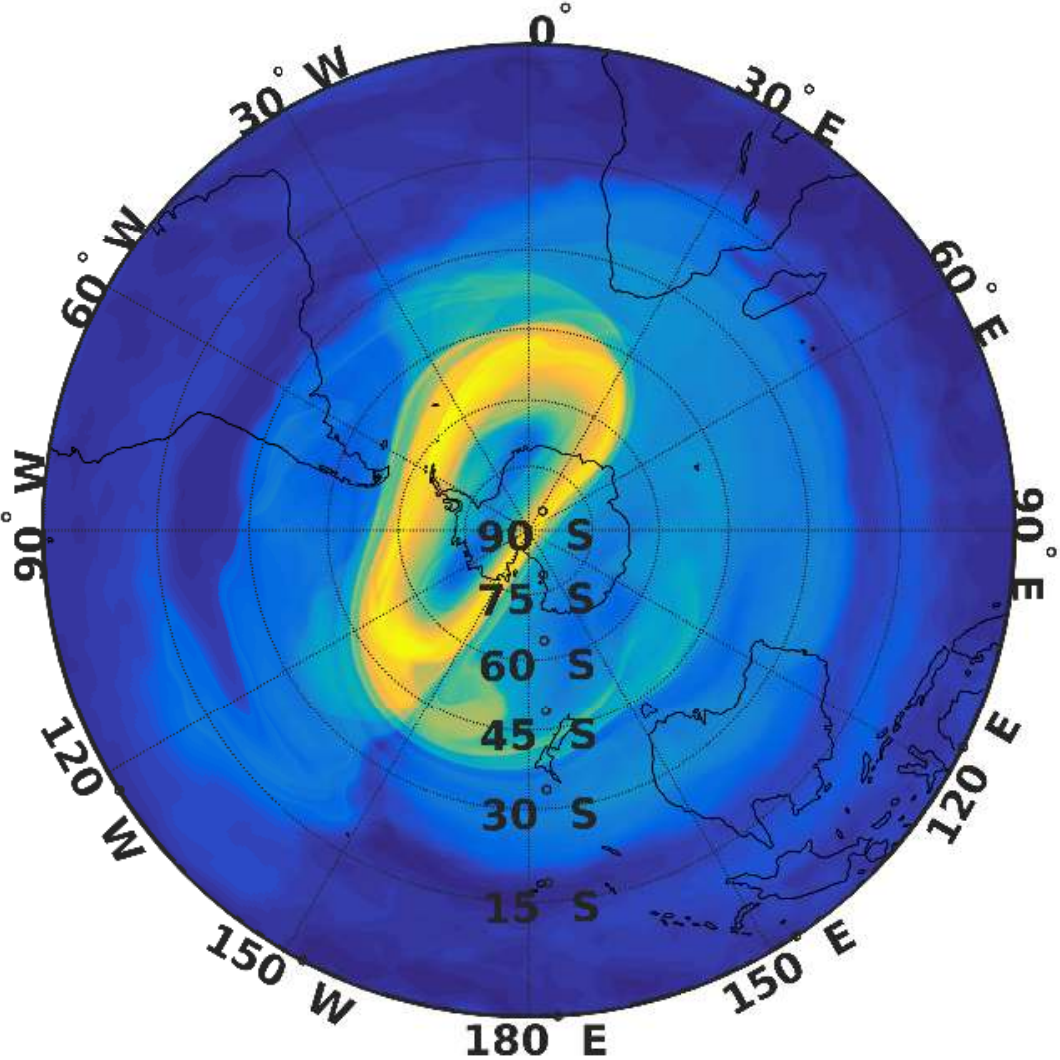}&  
\includegraphics[scale=0.3]{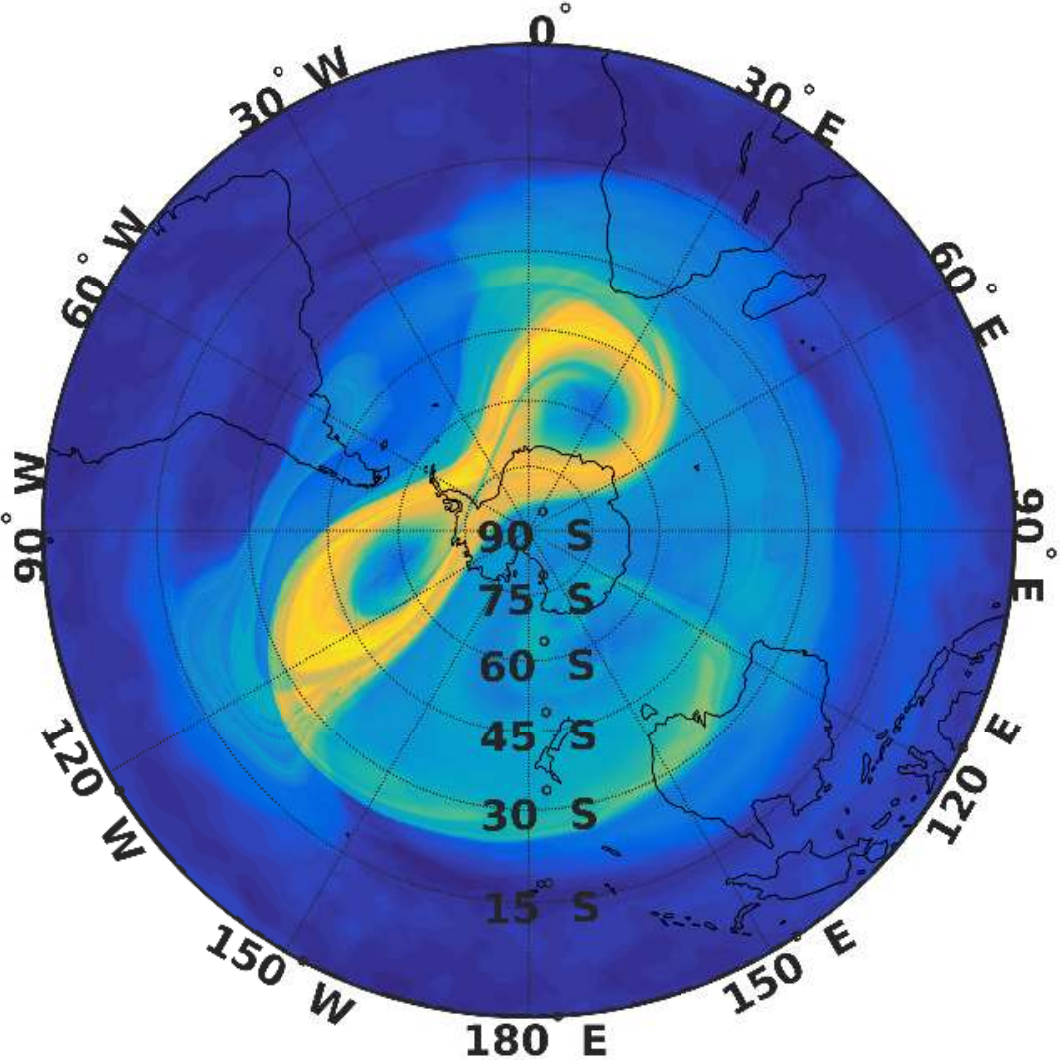}&\includegraphics[scale=0.3]{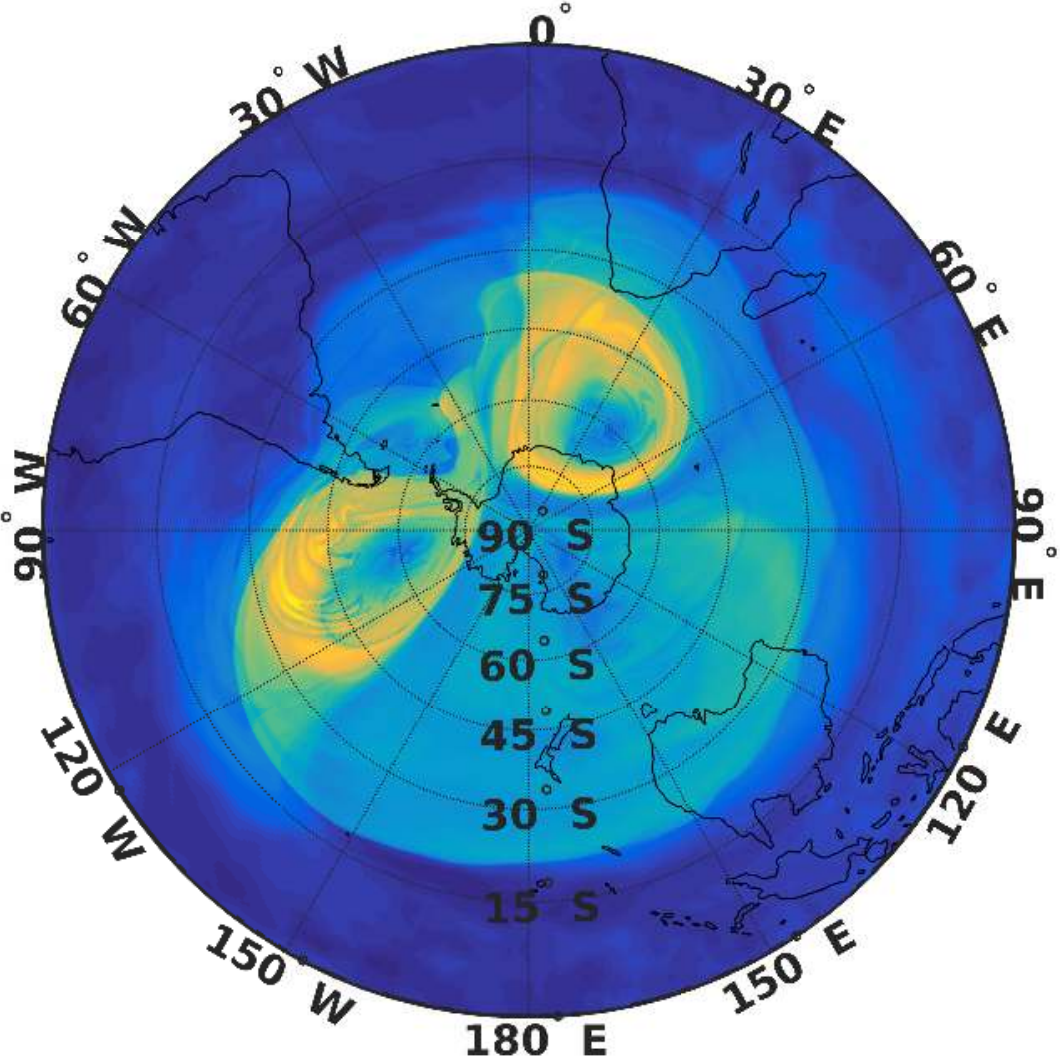}\\
  \rotatebox{90}{\hspace{2cm} 16 km}& 
 \includegraphics[scale=0.3]{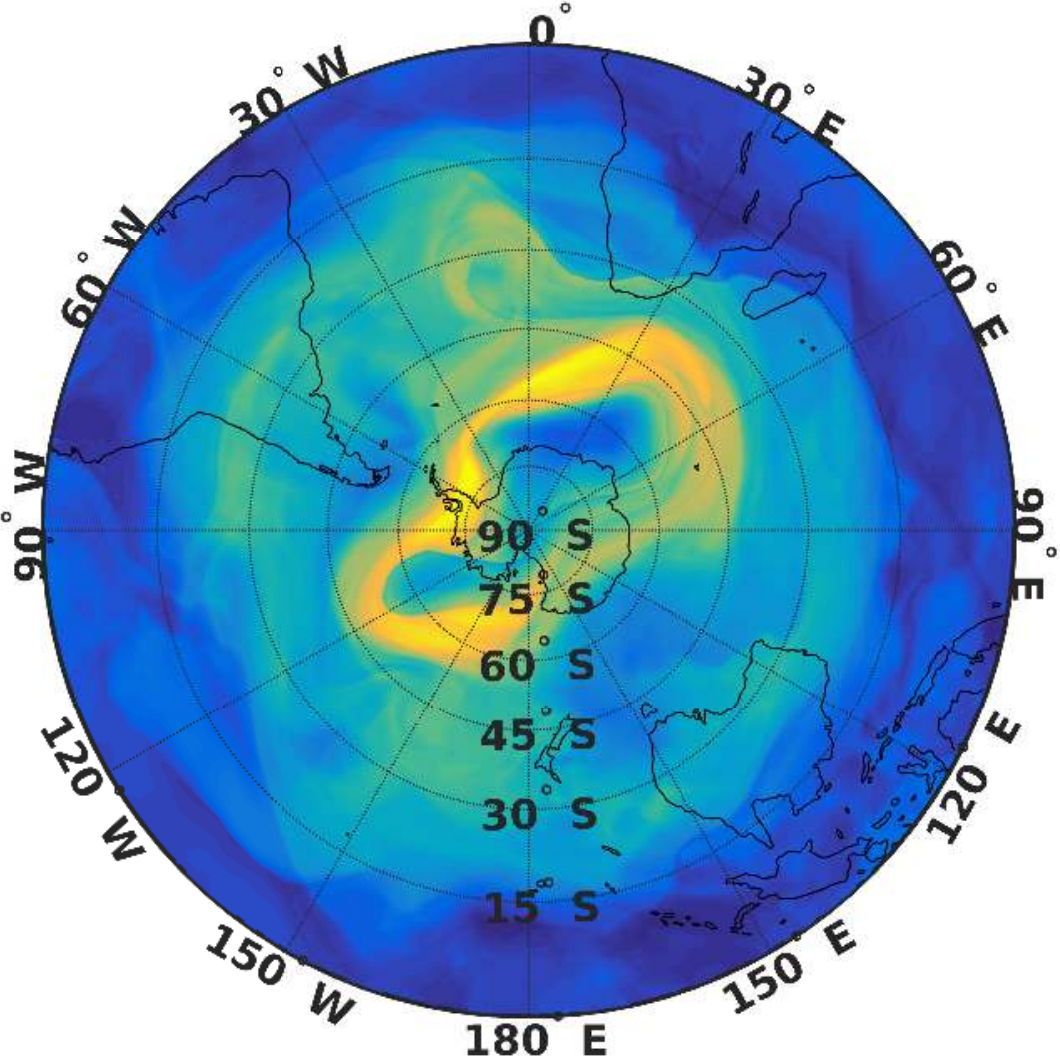}& 
\includegraphics[scale=0.3]{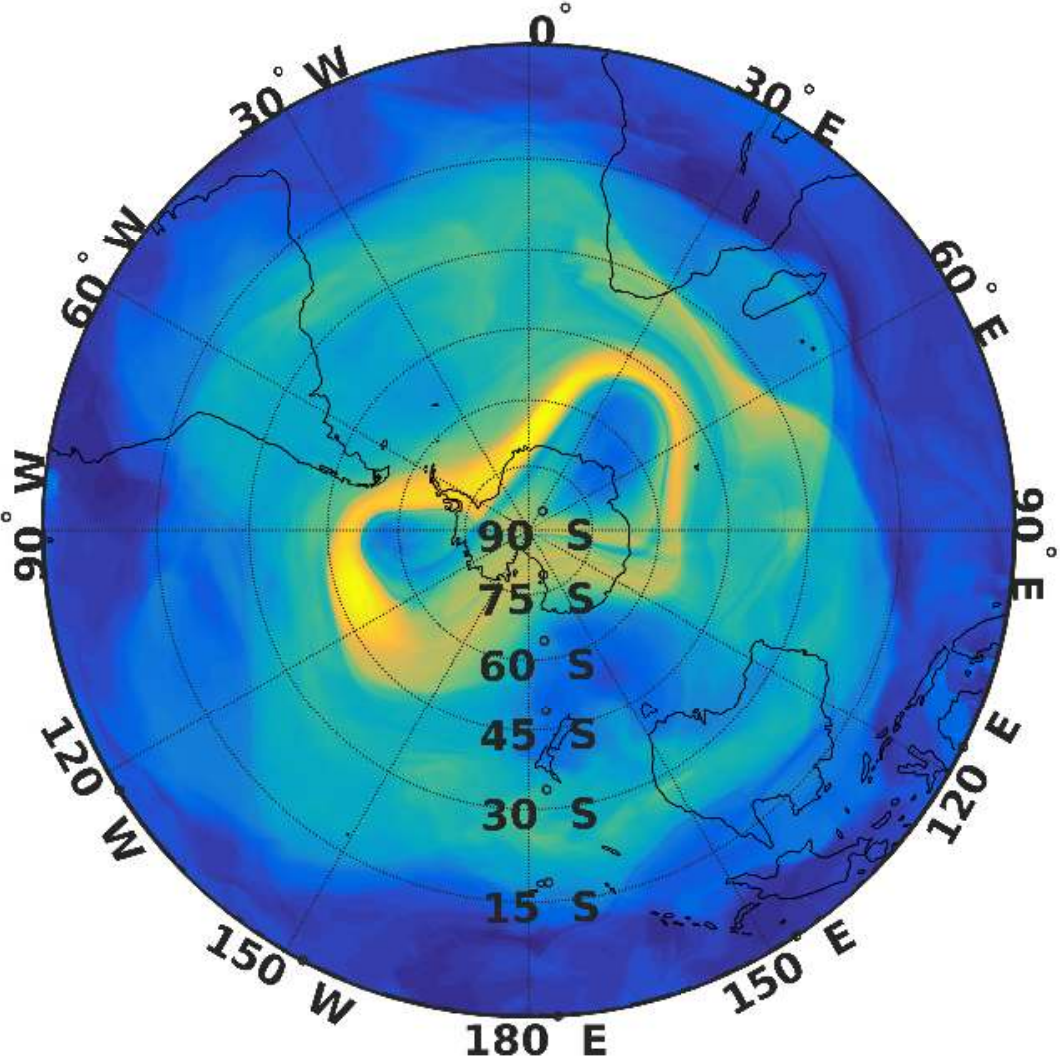}&\includegraphics[scale=0.3]{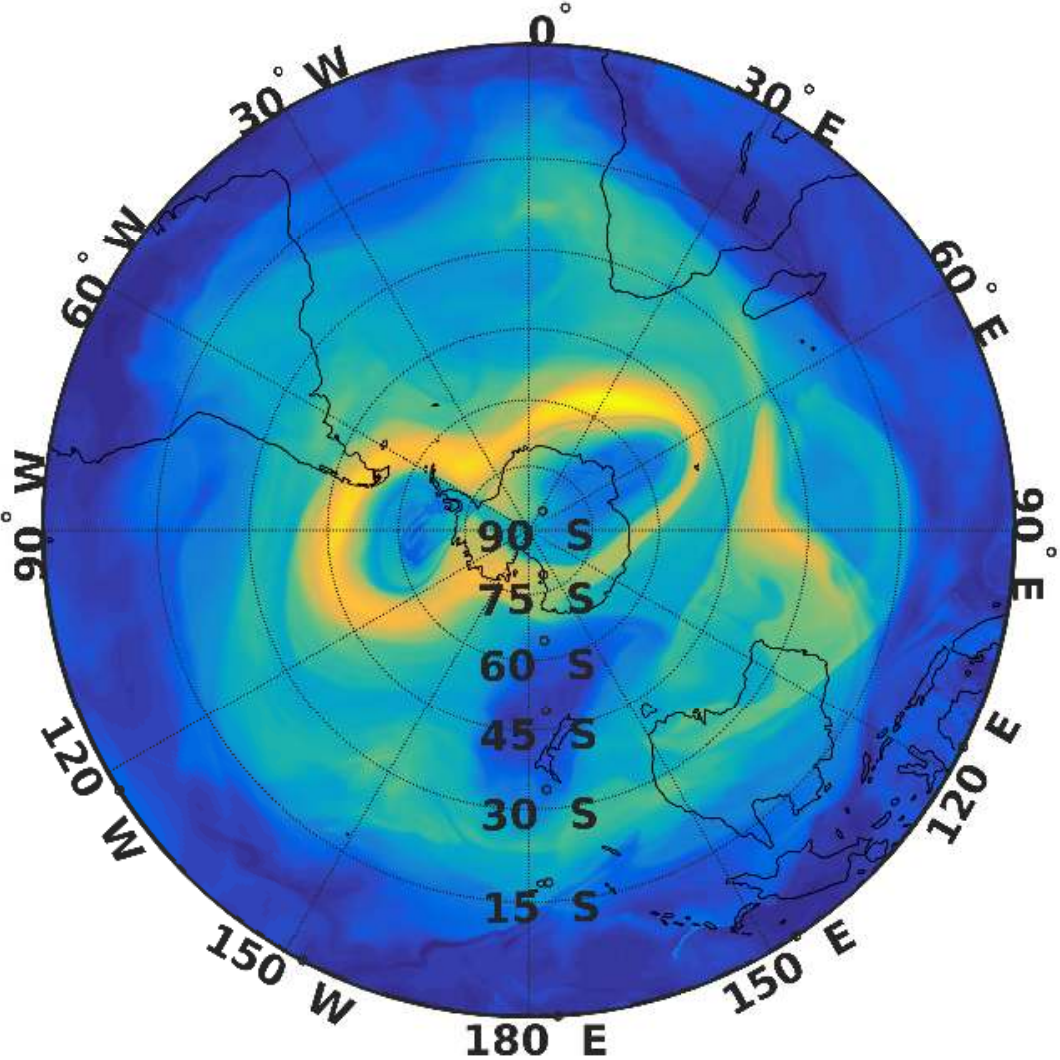}\\
   \rotatebox{90}{\hspace{2cm} 14 km}& 
 \includegraphics[scale=0.3]{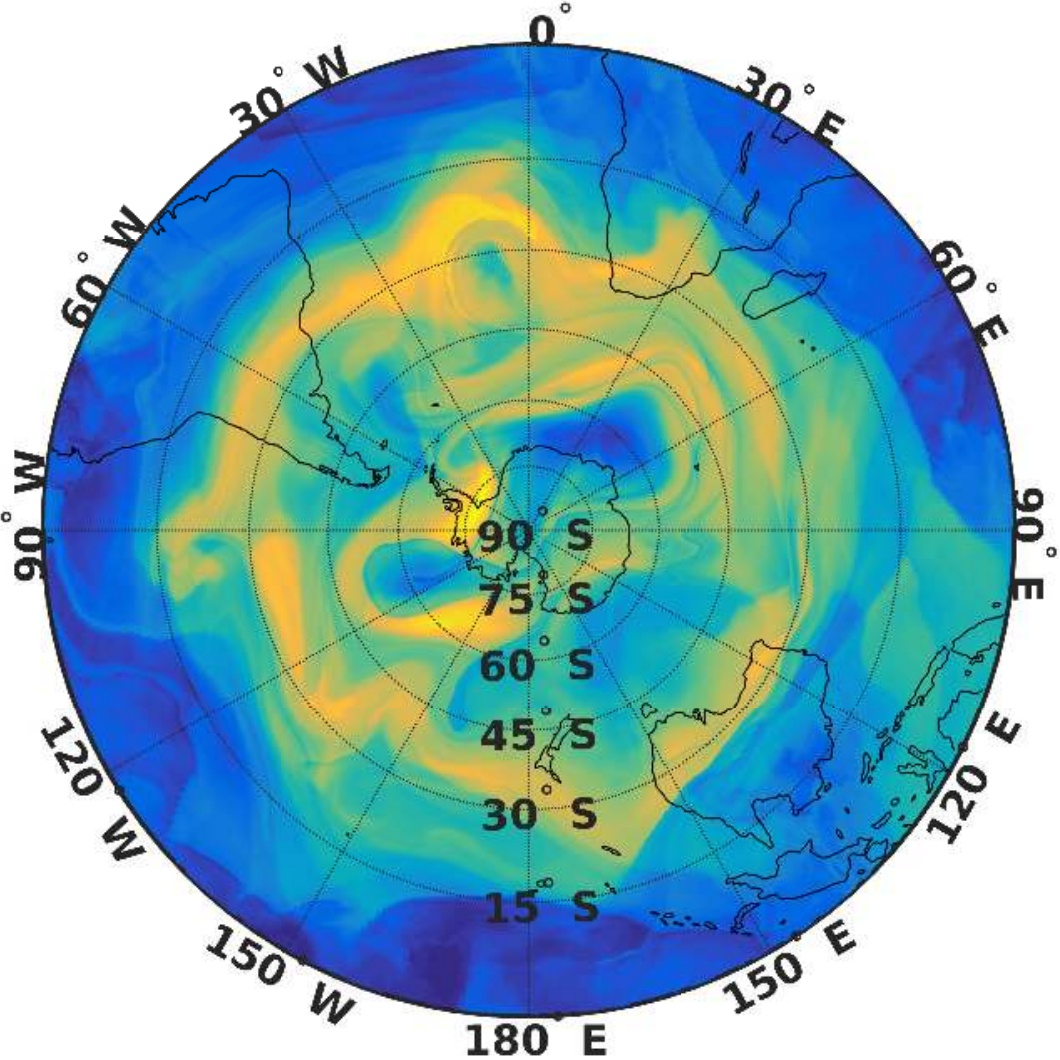}&  
\includegraphics[scale=0.3]{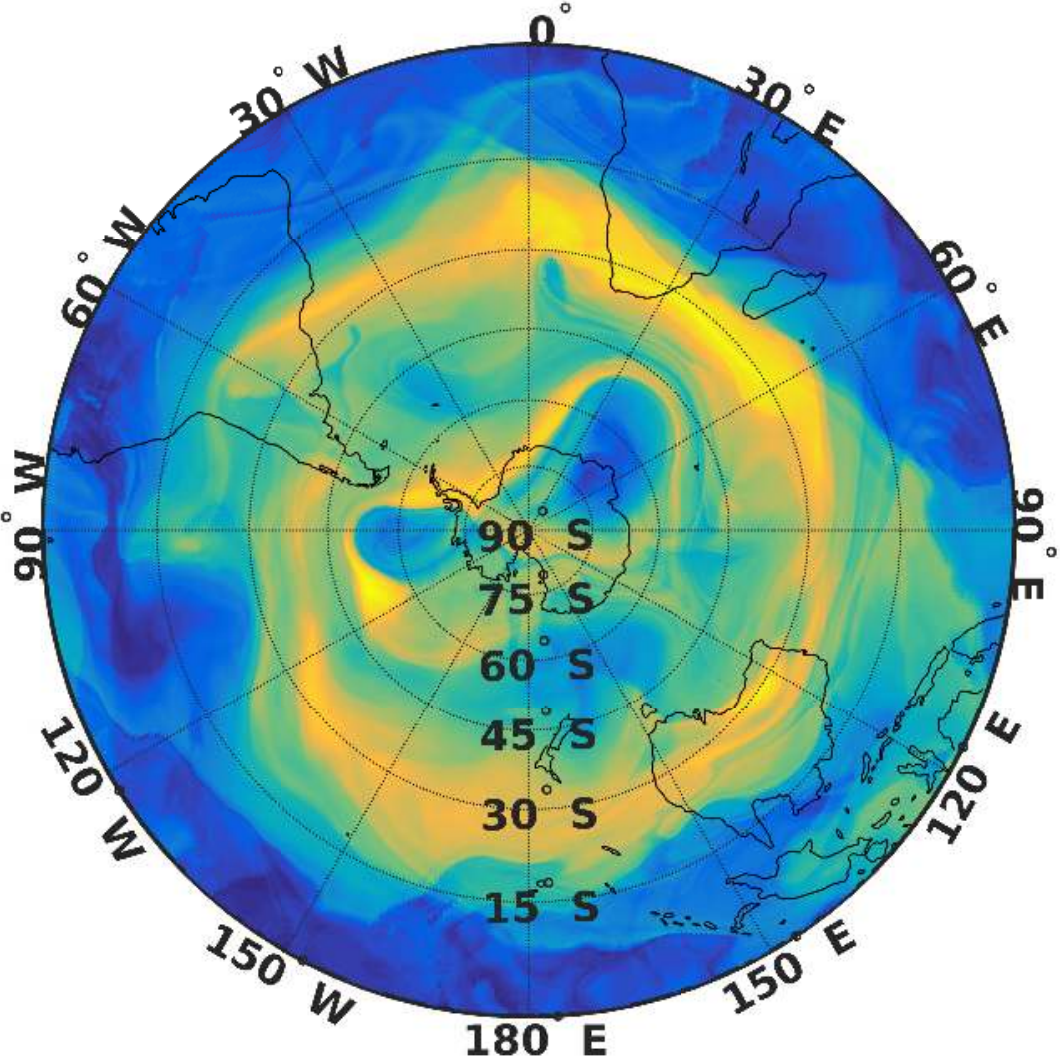}&\includegraphics[scale=0.3]{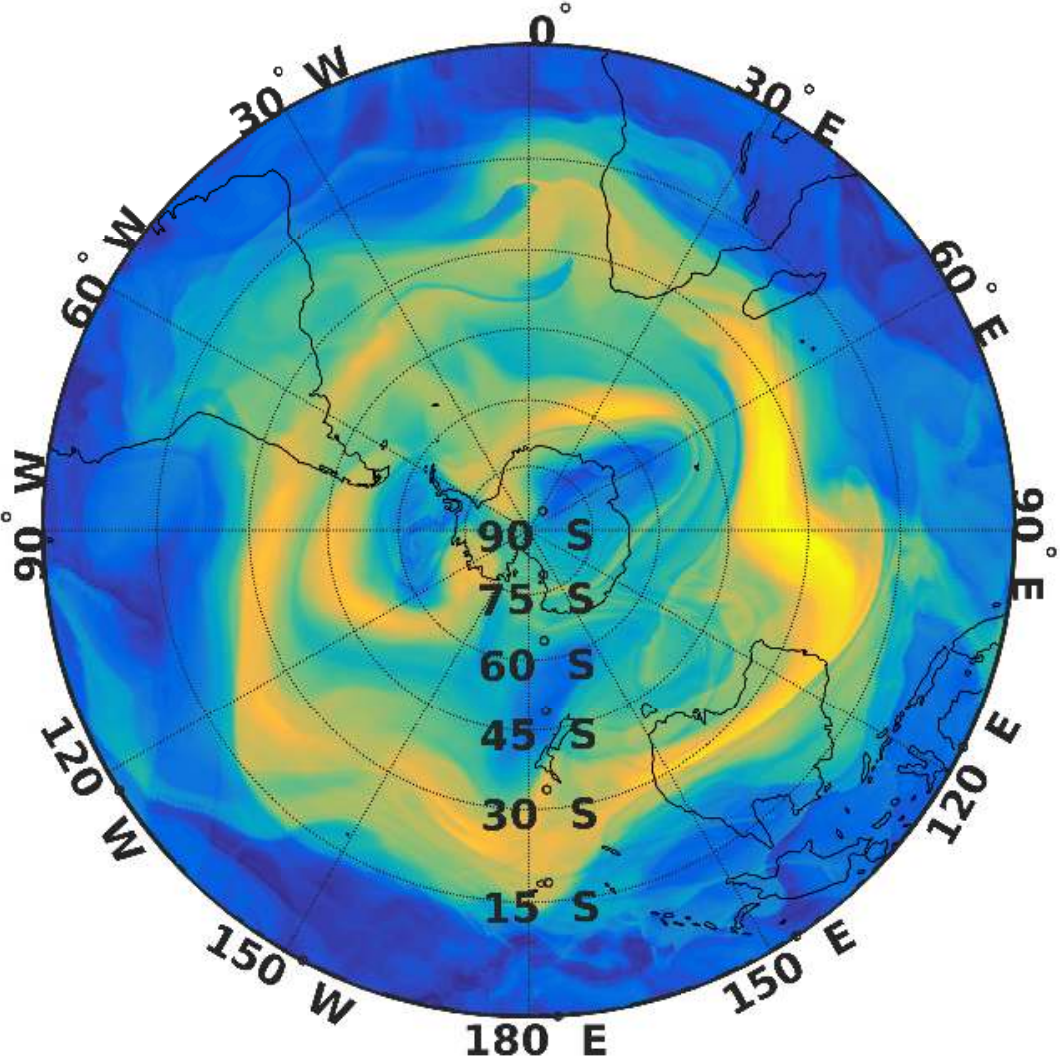}\\
   \rotatebox{90}{\hspace{2cm} 12 km}& 
 \includegraphics[scale=0.3]{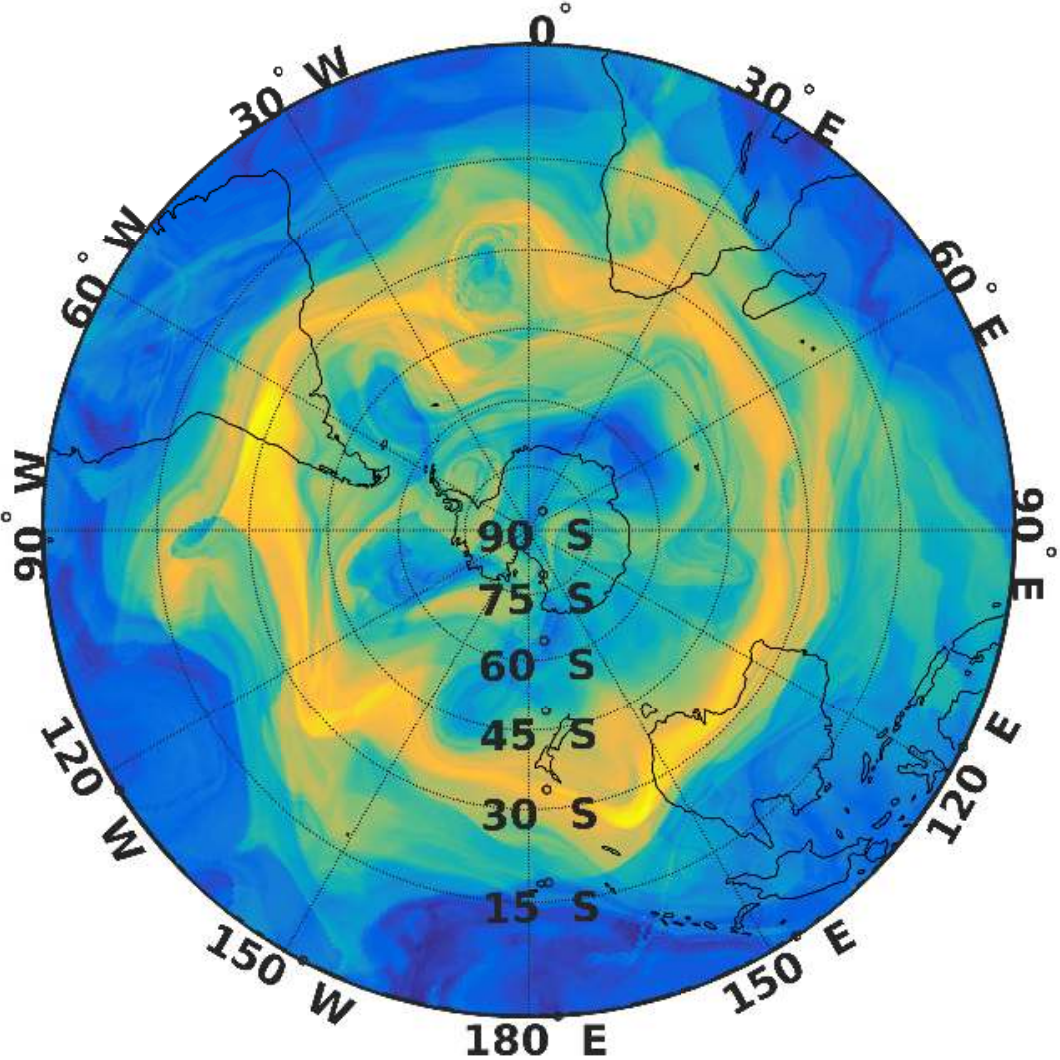}& 
\includegraphics[scale=0.3]{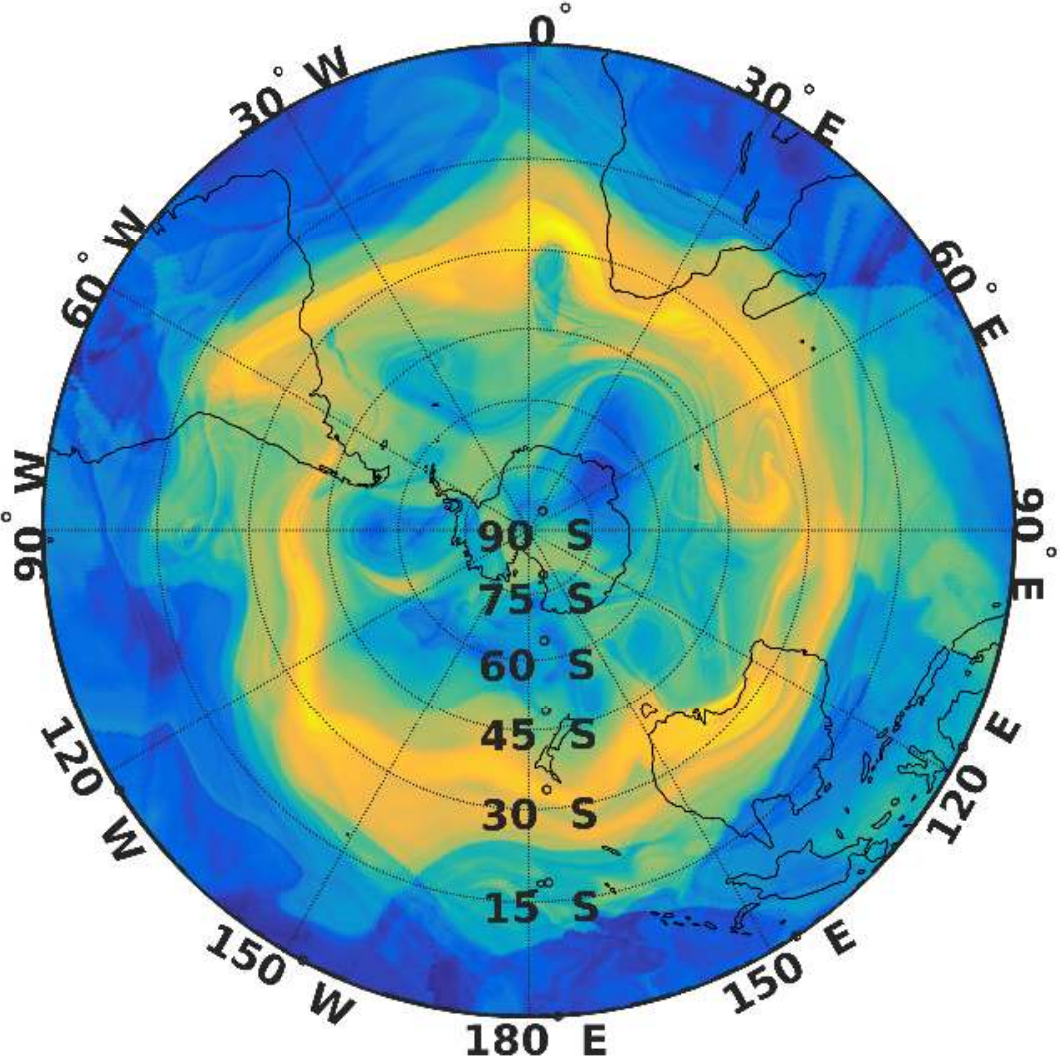}&\includegraphics[scale=0.3]{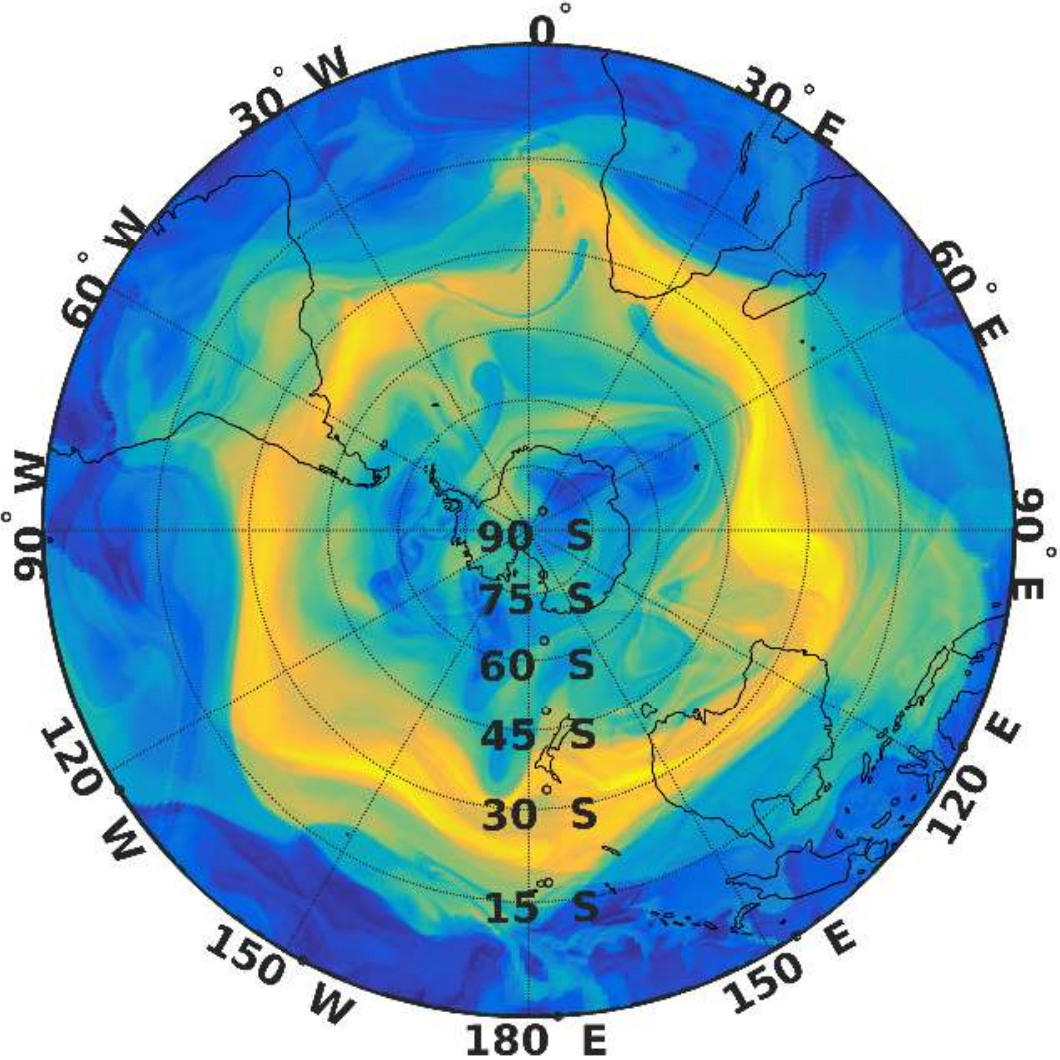}
 \end{tabular}
 \centering
\includegraphics[scale=0.5]{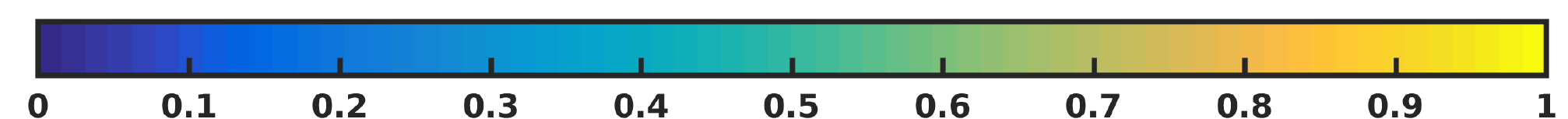}
 \end{center}
\caption{Multilevel plots (12, 14, 16, 30, 40 km) of the function $M$ ($\tau=5$) for 22 September (left column), 24 September (middle column), and 26 September 
2002 (right column). The values of $M$ are normalized by the maximum at each level and at the time of the plot.The largest and smallest values of $M$ in all 
figures are indicated with bright yellow and dark blue colors, respectively.} 
\label{fig:several_z}
\end{figure*}

\section{Introduction}
\label{intro}

{Advances in data gathering and processing systems have allowed for the assembly of a pictorial view in three dimensions (3D), including air motions and 
composition, during stratospheric sudden warmings. Notably, \cite{Matthewman09} (see also references therein) used the European Centre 
for 
Medium-Range Weather Forecasts (ECMWF) Re-Analysis (ERA-40) to investigate the evolution of the observed major midwinter stratospheric sudden warmings in the 
northern Hemisphere for the period 1957-2002. They considered separately vortex-displacement and vortex-splitting events, and documented the differences 
between 
their evolving vertical structures. \cite{Butler2017} used data from 
six different reanalysis products to produce a sudden stratospheric warming compendium 
for a region that extends from the surface to the stratosphere, as the importance of stratospheric-tropospheric connections have become increasingly apparent. 

Evidence from 3D numerical simulations with high resolution points to the existence of complex flow structures during the warmings. For example, 
\cite{manney1994} simulated the evolution in the stratosphere of the event during February 1979 with a 3D primitive equation model of the stratosphere.  They 
found that strong vertical velocities can develop both in 
the lower 
and upper stratosphere during the events. Most studies such as those referenced above have been performed in a \emph{Eulerian} context. Progress in dynamical 
systems, especially over the last 15 years, is offering the possibility of studying complex 3D flows following a \emph{Lagrangian} approach. This is 
particularly appropriate for the stratosphere, where transport is of paramount importance.}

The {present two-part paper follows a  Lagrangian approach to study the} stratospheric polar vortex (SPV) in the southern hemisphere during the final 
warming in spring of 2002, {when  it} experienced a unique splitting at upper levels in 
late September.   In Part I we present our methodology and focus on the description of the processes at work for filamentation and vortex splitting on an 
isentropic surface in the middle stratosphere. Part I also includes an Annex with a concise review of the Lagrangian concepts we use. In this Part II we 
examine 
the three-dimensional (3D) evolution of the event with special emphasis on (i) vortex splitting, and (ii) formation of barriers to transport  in the 3D flow as 
introduced by \cite{JC17} (referred to as JC17). 

The analysis reveals a SPV over the polar region with a columnar two-lobe structure extending and branching unevenly upwards, and a distinct subtropical jet 
stream in the troposphere. The role of Lagrangian structures in determining the fate of parcels in reference to their organization in either one or two 
vortices 
is discussed in detail. We find compelling evidence of deep 3D barriers to transport in the stratosphere that from  the mathematical point of view can be 
identified with the Normally  Hyperbolic Invariant Manifolds (NHIM) described in JC17.

\begin{figure}
\includegraphics[scale=0.35]{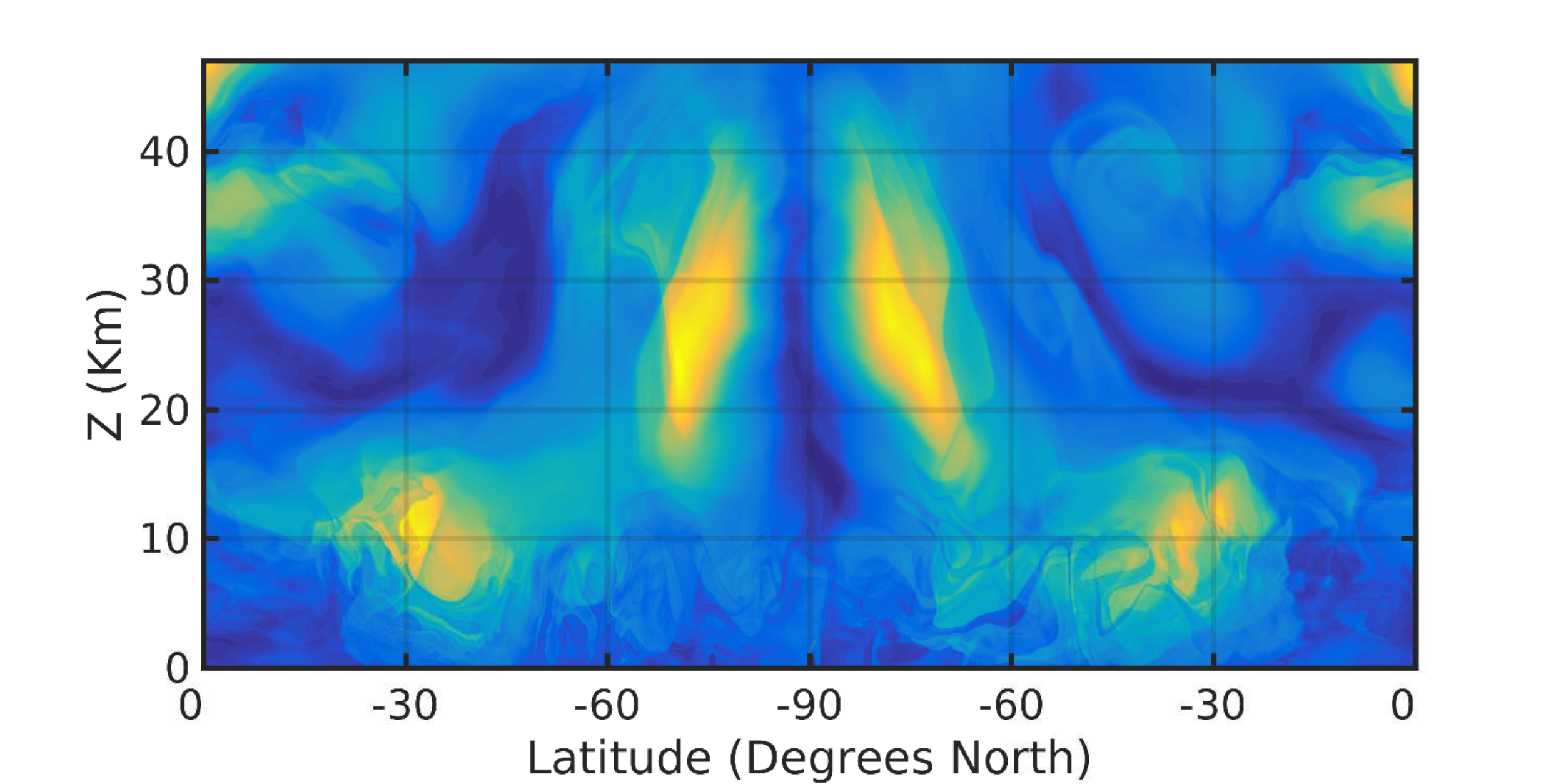}
 \centering
\includegraphics[scale=0.37]{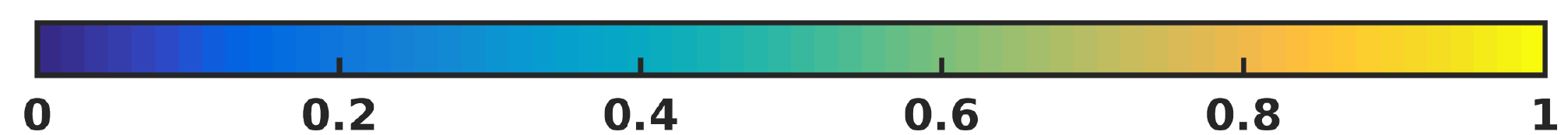}
 \caption{Vertical cross section of the function $M$ ($\tau=5$) on 15 October 2002 at $90^\circ$W and $90^\circ$E. Both the subtropical jet and the SPV are 
highlighted by the yellowish features.} \label{fig:15october}
\end{figure}

Our principal analysis tool is the Lagrangian descriptor known as the function $M$.  This is defined in Part I, where we review its properties in the context 
of 
2D flows. In Part II we also use the ERA-Interim reanalysis dataset produced by the European Centre 
for Medium-Range Weather Forecasts (ECMWF; \cite{simmons07}).

JC17 presents a methodology to compute $M$ from 3D velocity fields in the ERA Interim reanalysis. {The first step consists of obtaining the 
parcel trajectories.  A special feature of our calculation procedure is that, on z-constant surfaces and in order to bypass singularities at the pole with the 
spherical coordinates, parcels are advanced on a cartesian coordinate system with a Runge-Kutta Method  that used
a time step of 1 hour. The vertical velocity $w$ $(m/s)$ used to advance parcels in the vertical is calculated from 
$\omega$ (Pa/s), 
temperature and specific humidity provided by ERA-Interim. Once the trajectories are computed, M is obtained on a spatial grid of 600x500 points. Next, the 
principal issue is the interpretation of the Lagrangian descriptor features in terms of  hyperbolic trajectories and their invariant manifolds.  This 
interpretation issue is not fully resolved at the moment for the 3D context because it requires  theoretical guidance that needs further development. Some 
help in our case is provided by the fact that on appropriate time scales, stratospheric flows are quasi 2D in the sense that the magnitude of the vertical 
velocity component is much smaller than the horizontal velocity. For such flows, JC17 exploited the concept of normally hyperbolic invariant manifold (NHIM), 
which we will use in this Part II. }

\begin{figure}
\begin{tabular}{ll}
(a) $M$ on 24 September 2002 00:00:00  & (b) $M$ on 24 September 2002 18:00:00\\ 
 \includegraphics[scale=0.7]{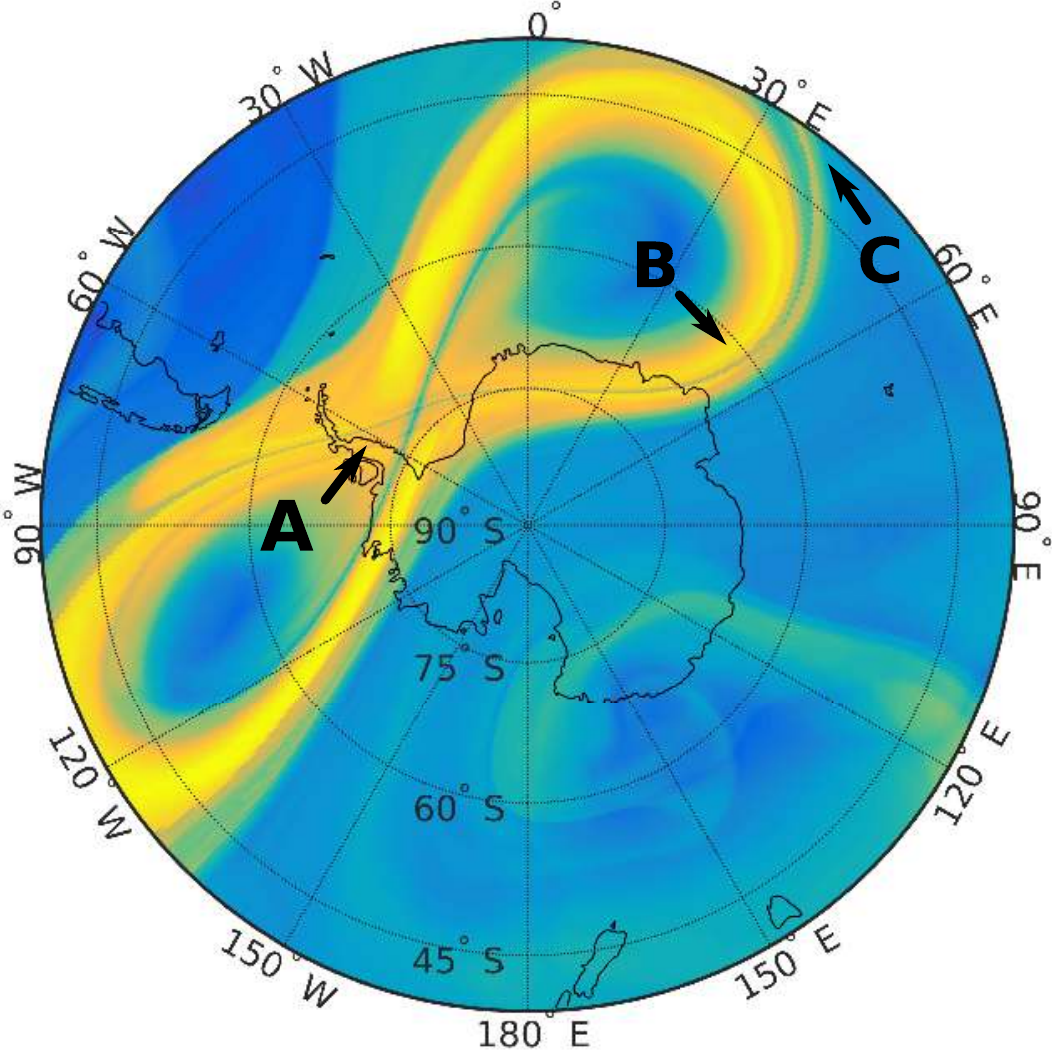} & 
\includegraphics[scale=0.7]{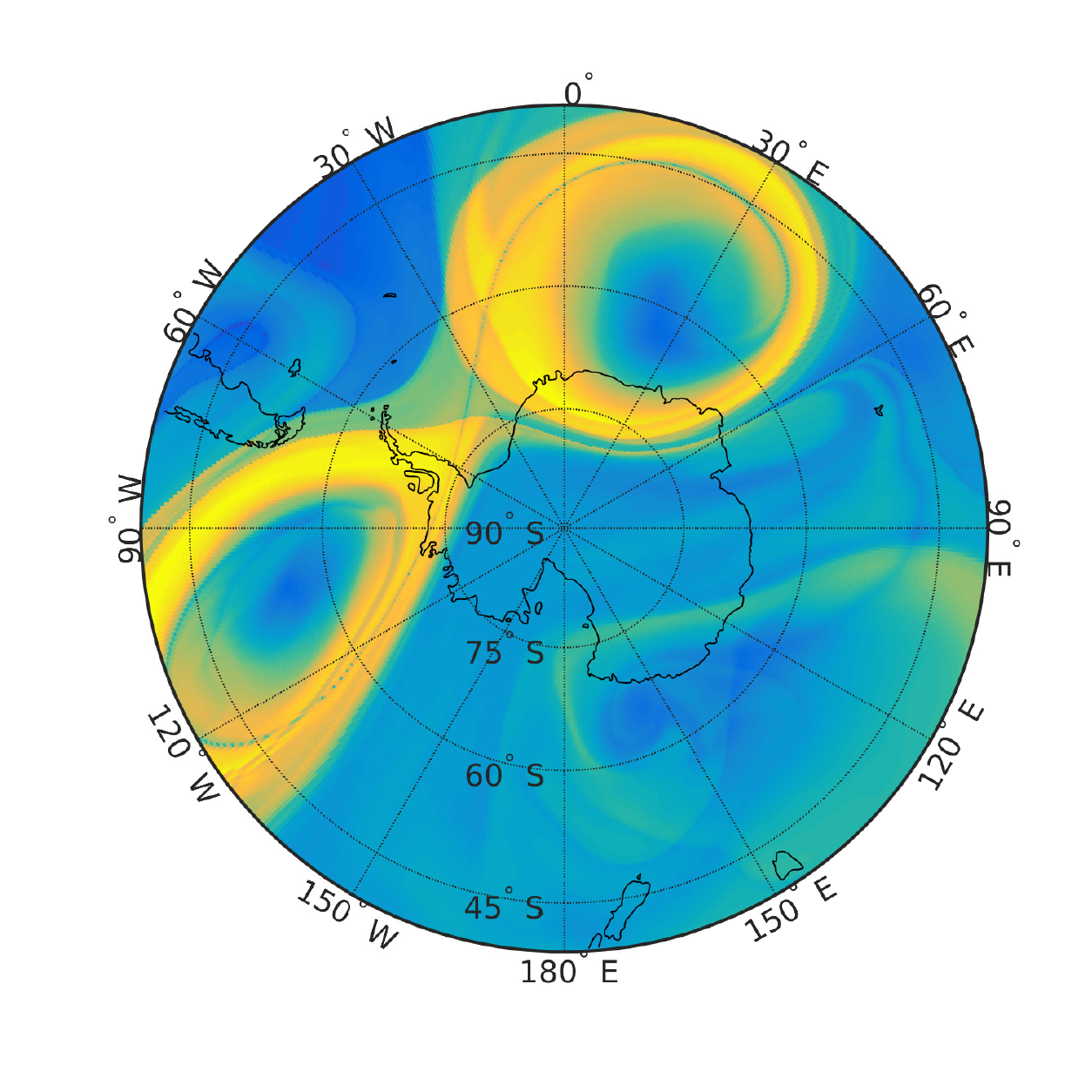}\\
\multicolumn{2}{c}{\includegraphics[scale=0.5]{colormap_M_padula.pdf}}\\
(c) $\nabla M$ on 24 September 2002 00:00:00  & (d) $\nabla M$ on 24 September 2002 18:00:00\\
 \includegraphics[scale=0.7]{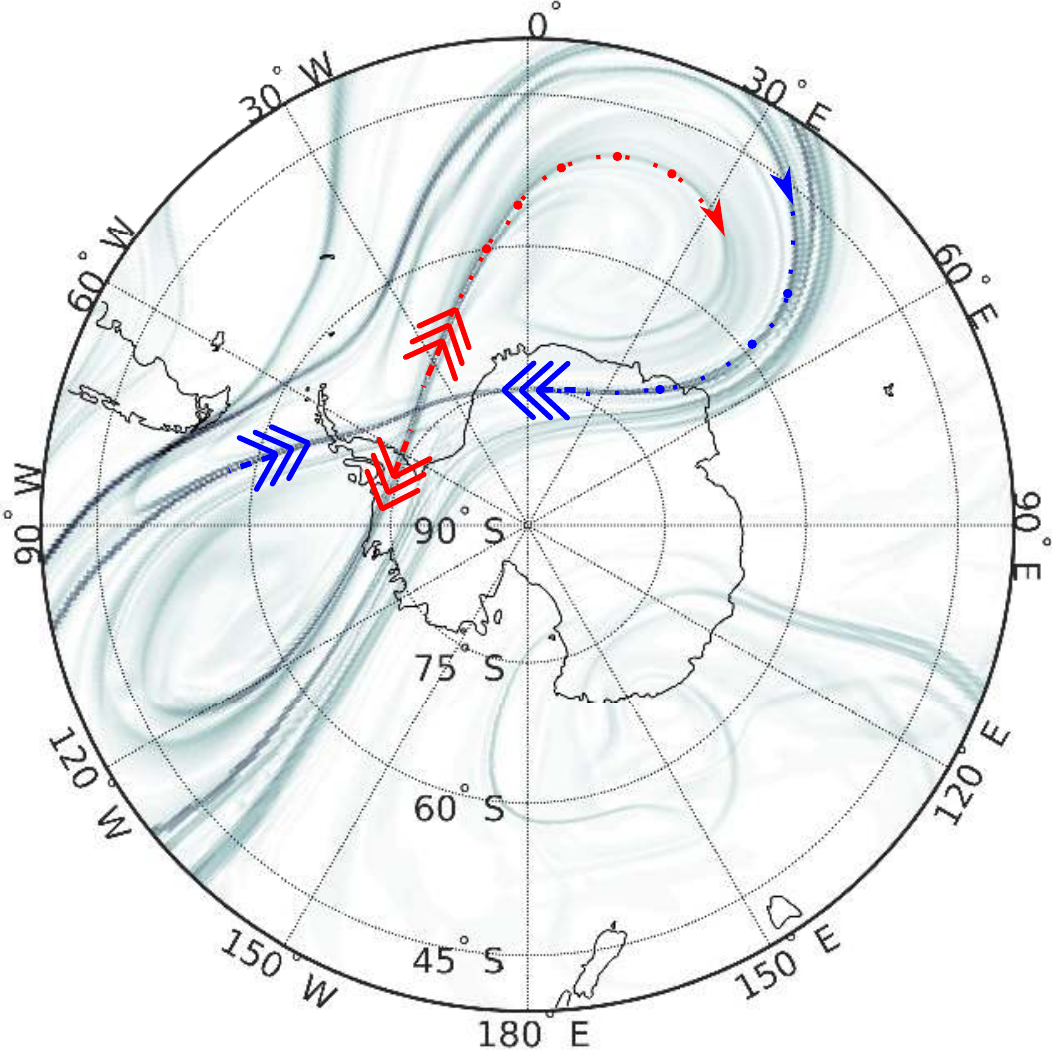} & 
\includegraphics[scale=0.7]{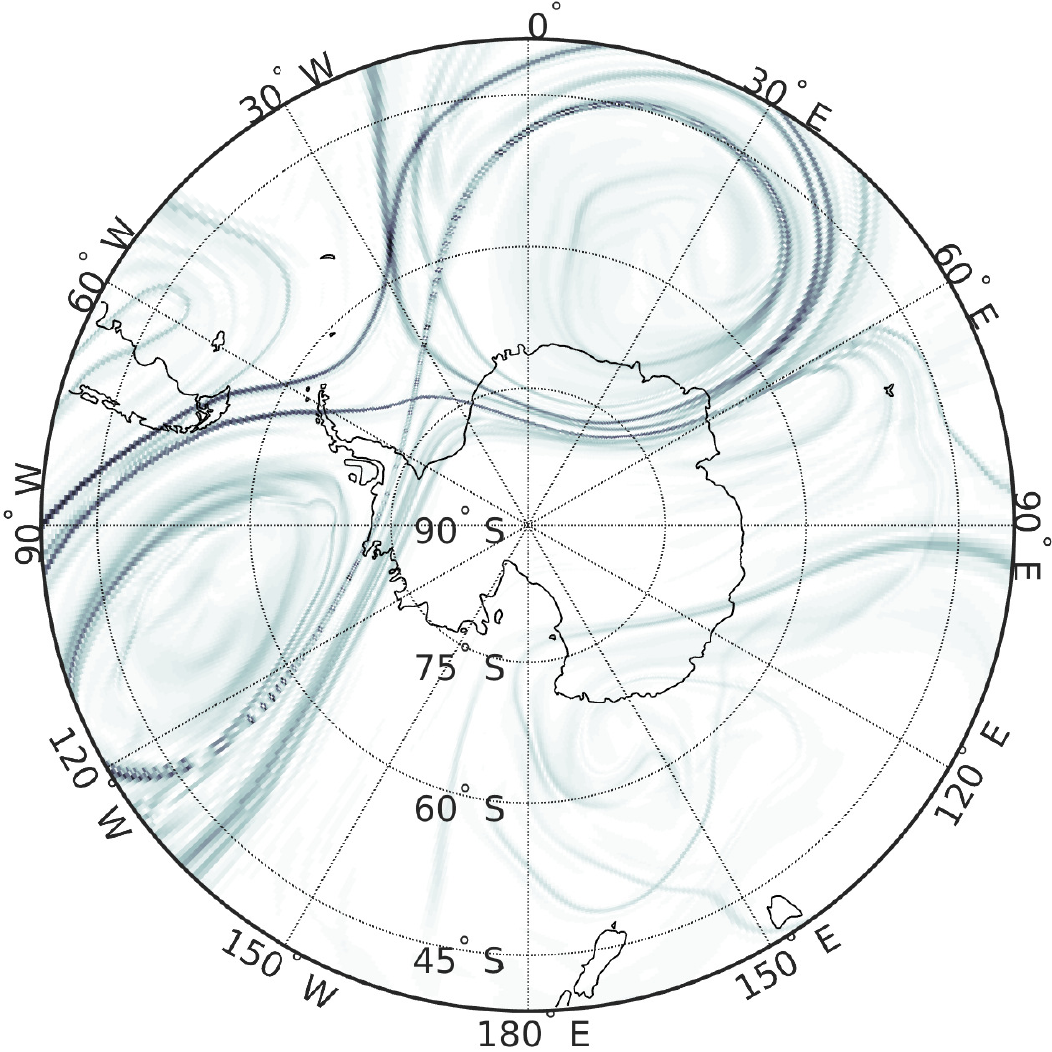}\\
 \end{tabular}
 \caption{The function $M$ ($\tau= 5$) at 850K shortly before (a) and after (b) the vortex split at that isentropic level. The left panel shows a stable and an 
unstable manifold, which are highlighted by blue and red arrows, respectively. The meaning of the labels  A, B, C is explained in the text.}\label{fig:z31}
\end{figure}

We give a simple illustration of the NHIM concept as follows.  Let us consider the 2D flow in the neighborhood of a hyperbolic point (x$_0$, y$_0$) fixed in 
time together with its associated unstable and stable manifolds.  A flow in the 3D space can be defined by ``stacking up'' the 2D surfaces in the z direction.  
In this flow, the hyperbolic points form a line in the z direction and the manifolds form vertical surfaces and act as 2D barriers to the particles.  This is 
an idealization of the NHIM concept. Since NHIMs persist under perturbation (i.e. both horizontal and vertical perturbations)  \cite{Wiggins_book1994, 
Mezic1994},
the application of this concept to stratospheric flows is relevant since the vertical component of velocity is significantly smaller than the horizontal 
component. Consequently this approach allows us to conclude the existence of similar structures in the reanalysis data. { Hints of the NHIMs are given in the 
paper by \cite{Schoeberl95}, who performed a trajectory analysis of a large 
wavebreaking event in the northern polar vortex during 1992-1994 and found that such filaments have a deep vertical structure.}

{The recent paper by \cite{RCD2018} represents an advance in the interpretation of the NHIM and its structures in 3D vector fields for which analytic solutions 
are known or can be obtained by
exact calculations.  The results  of this paper, as well as others currently in preparation, give us confidence in working with such structures as detected by 
our Lagrangian descriptor.} 

\begin{figure*}
\begin{tabular}{ll}
(a) 9 September 2002 00:00:00  & (b) 24 September 2002 09:00:00\\
   \includegraphics[scale=0.7]{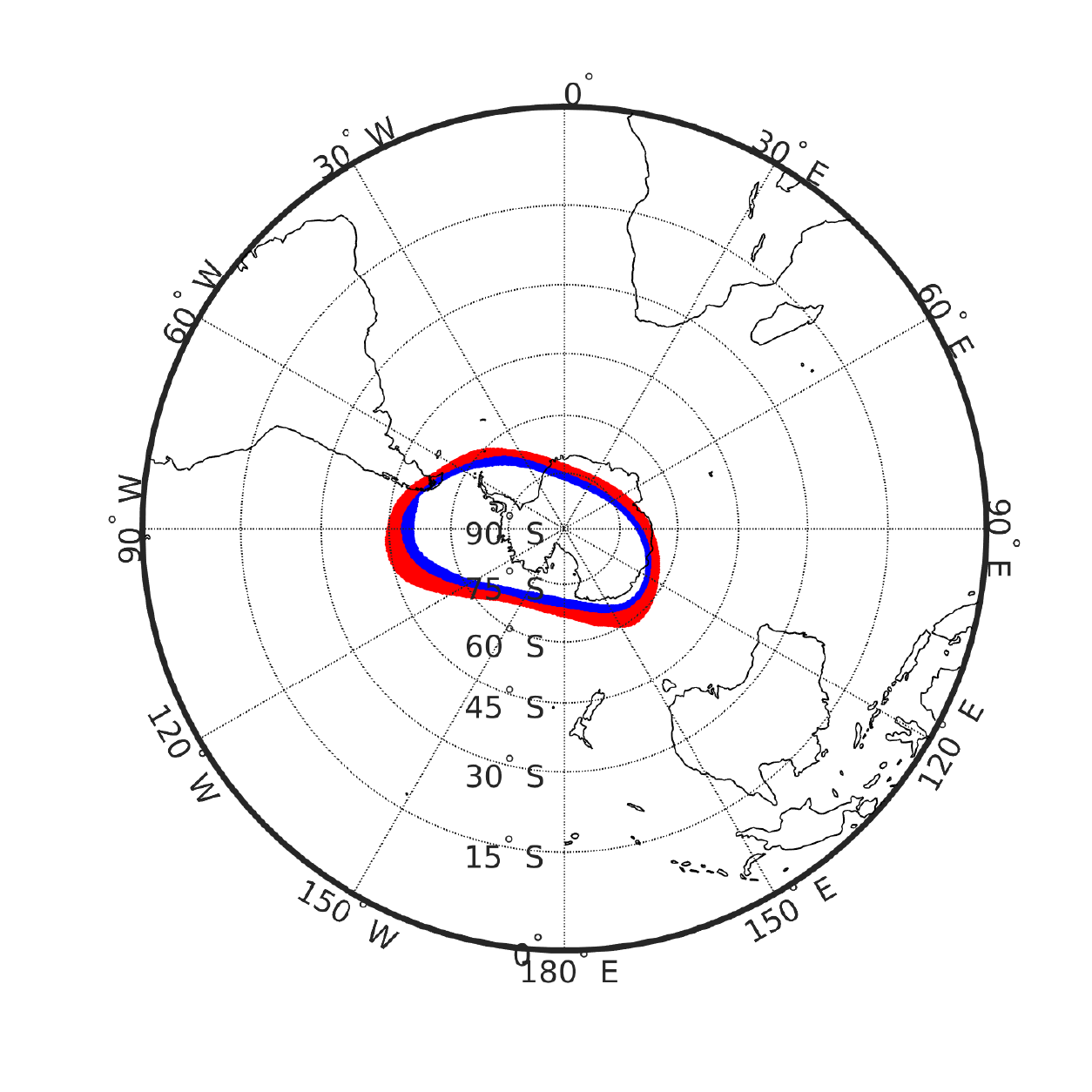}&
   \includegraphics[scale=0.7]{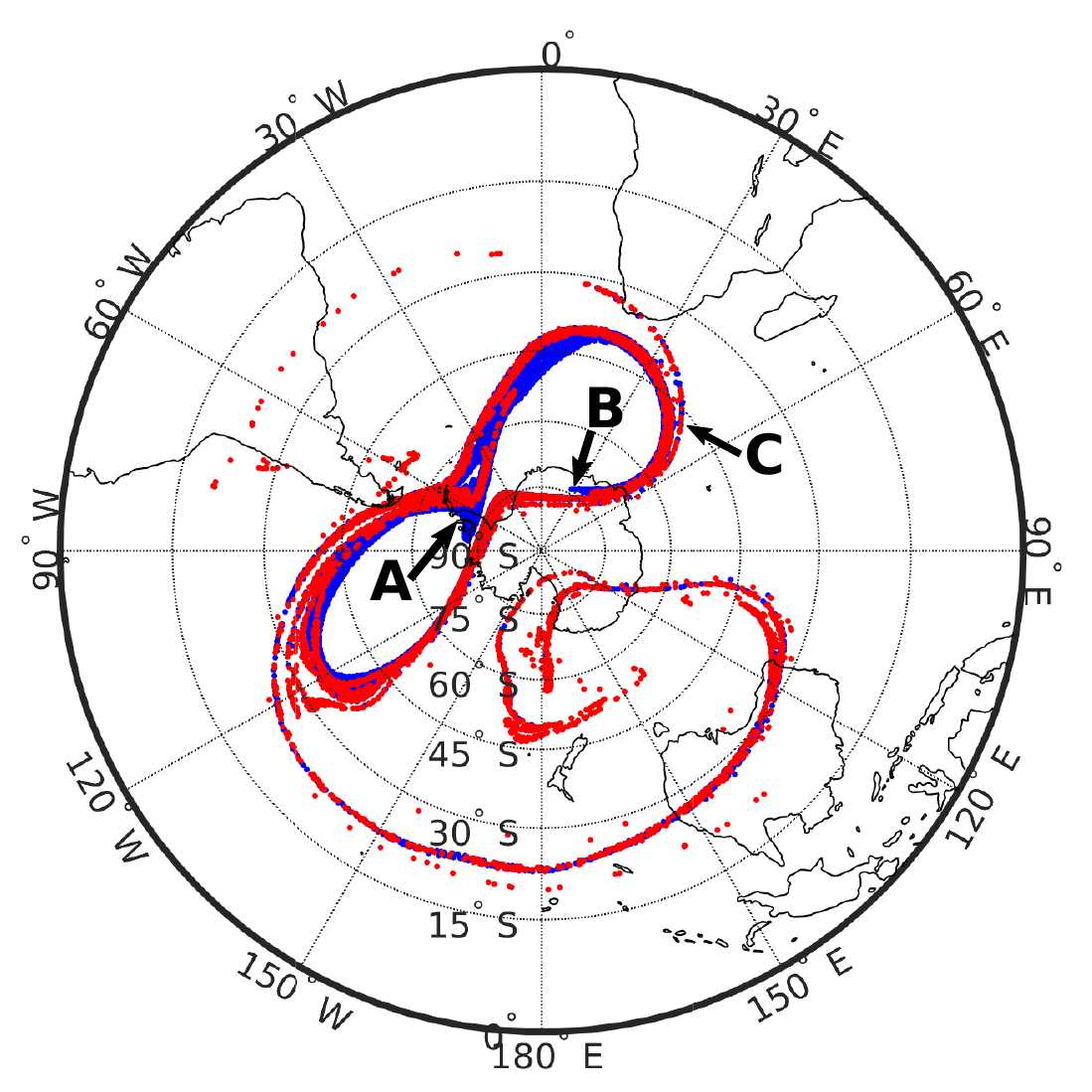}\\
  (c) 25 September 2002 01:00:00   & (d) 26 September 2002 05:00:00\\
   \includegraphics[scale=0.7]{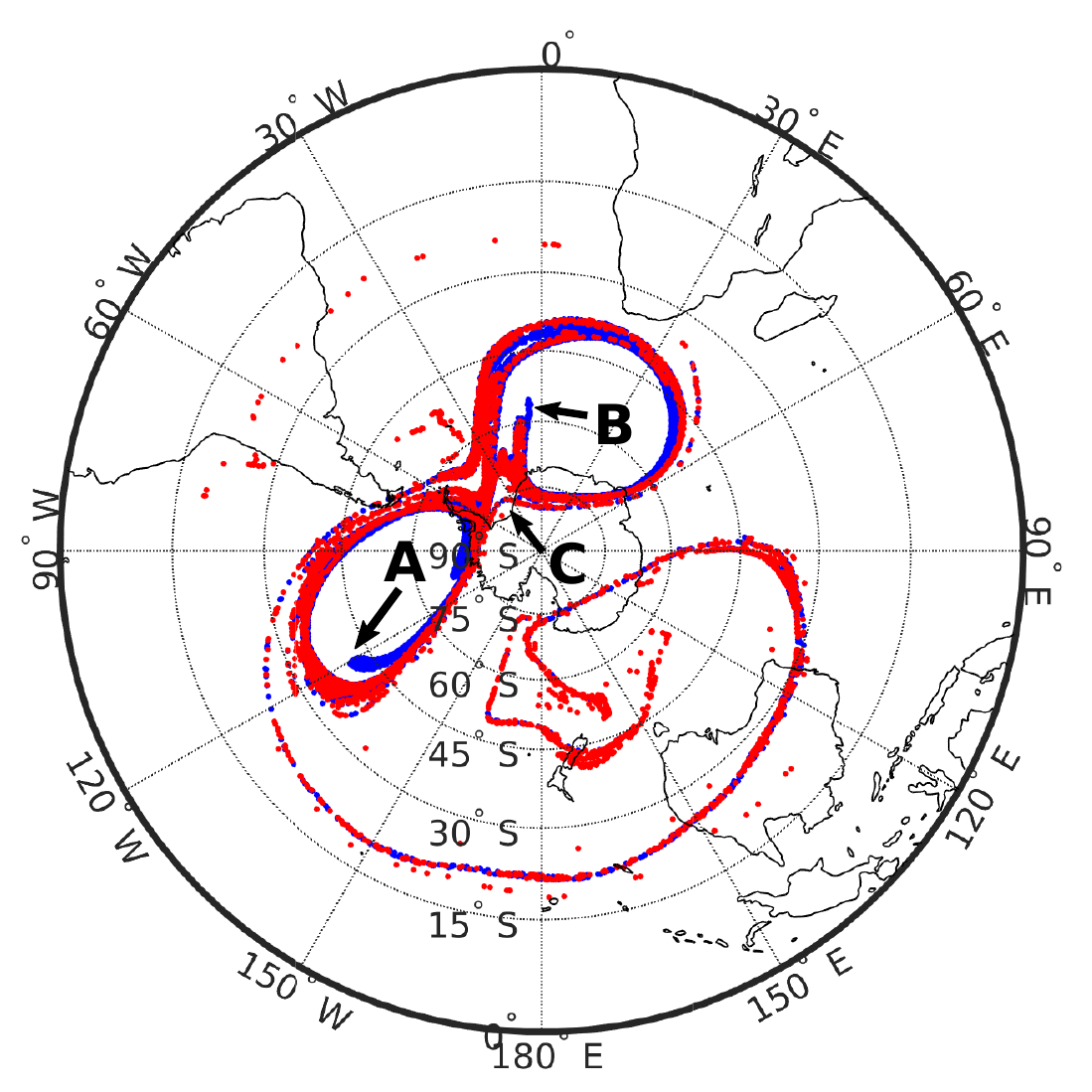}&
  \includegraphics[scale=0.7]{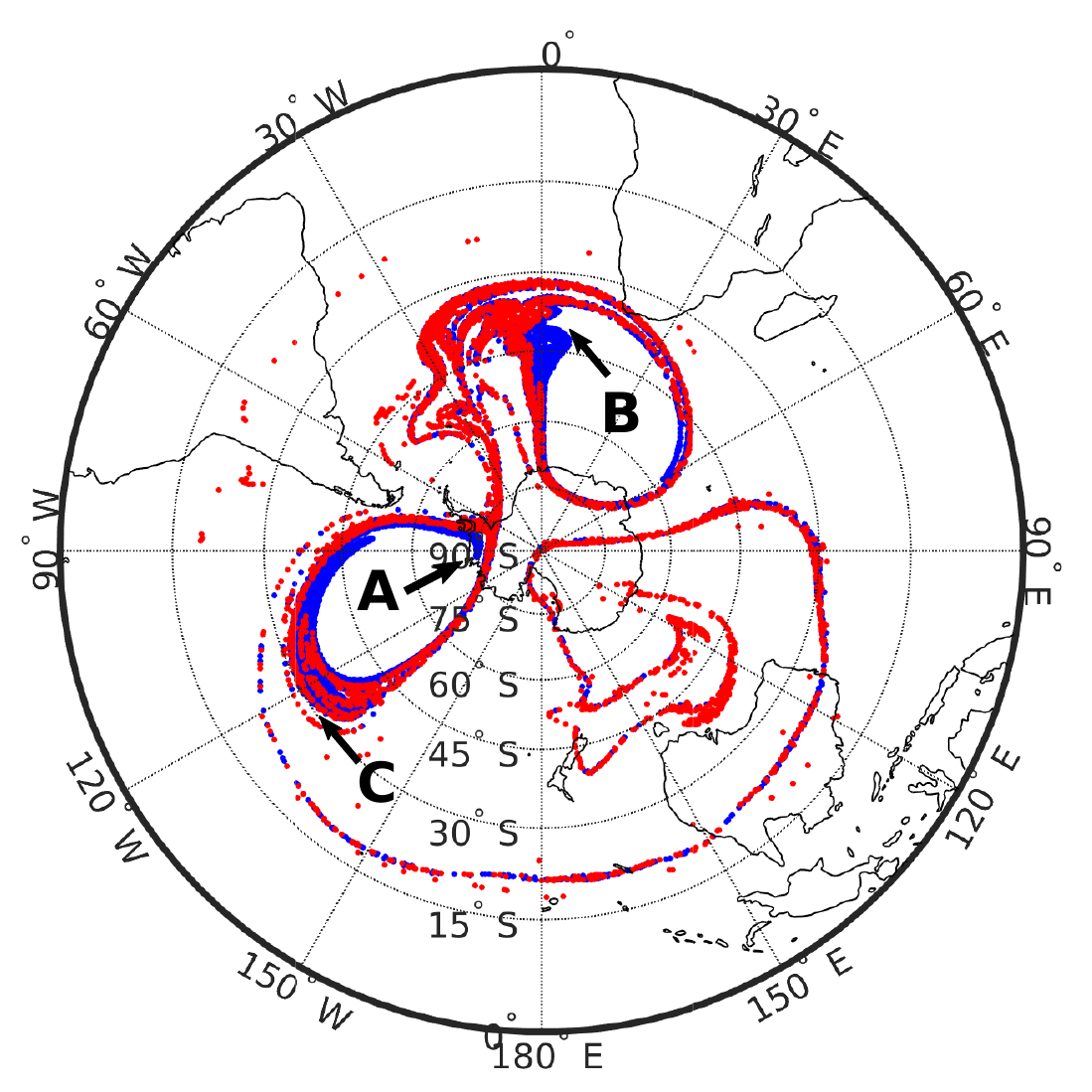}\\
  \end{tabular}
 \caption{{Consecutive position at 850K of the particles selected in (a) for 9 September 2002 at different times of the SPV splitting}.  All 
{selected} particles 
are between the contourlines corresponding to $M$ =$7.7 \cdot 10^4$. The 
particles  are drawn in  either blue or red to differentiate  those that are inside or outside the contour defined by the maximum value of $M$ at each 
longitude. The meaning of the labels A, B, and C is explained in the text.} 
\label{fig:particle2Ddata}
\end{figure*}

Part II is organized as follows.  
Section \ref{M3D} presents the Lagrangian descriptor at several constant height levels from the upper troposphere to the stratosphere.  Section \ref{split} 
details the particle evolution in the middle stratosphere during the SPV splitting. Section \ref{Vertical Structure} describes the vertical structure of the 
flow during the event.  Section \ref{s:summary} consists of a summary and a list of conclusions.

\section{The southern troposphere-stratosphere  in spring 2002}
\label{M3D}

The values of $M$ on 22, 24, and 26 September at different  levels in the vertical are presented in Figure  \ref{fig:several_z}. {The  figure 
corresponding to Figure  \ref{fig:several_z} at
constant potential temperature levels is in the supplementary material from which similarities are visible specially at upper 
levels. In the present paper, we use the z-coordinate when we wish to emphasize geometric structures in 3D, such as those are probed by remote sounding 
devices. We work in isentropic coordinates when our focus is on the behavior of trajectories in reference to the edge of the polar vortex.  This is done 
because 
the vertical displacements of parcels from horizontal surfaces can be much larger than from isentropic surface and hence trajectories in the latter system are 
more representative than in the former. In plots of $M$, bright yellow and dark blue colors are assigned to the highest and lowest values, respectively. } 

{Figure 1 shows a meandering yellow band in the high latitudes at and above 14 km.} This is the signature of the SPV. On 22 September, the SPV has started 
to pinch at 16 km. This process continues on 24 
September, and by 26 September the SPV has clearly split above 16 km and two distinct vortices have formed. Also at all times, we can see the 
signature of the subtropical jet stream at and below 14 km as another meandering, bright yellow band at around 30$^o$S. The jet stream break just west of South 
America resembles the 
Rossby wave breaking in the Double Downstream case of \cite{Peters2003}.  In such a case, air with low PV intrudes poleward along the western flank of South 
America, contributing to define a cyclonic region in the southeastern Pacific.

Figure \ref{fig:15october} shows the vertical Lagrangian structure of the troposphere and stratosphere once   the SPV partially recovered in the upper 
stratosphere after the splitting at the end of September.  The function $M$ has been computed with $\tau=5$ days on the 15 October 2002 on the vertical cross 
section $90^\circ$W and $90^\circ$E. The strongest values of $M$ correspond to the subtropical jet stream centered at about $30^\circ$S, 10 km height, and the 
SPV extending almost vertically at higher levels around the pole at about $70^\circ$S.

\section{The SPV split in September 2002}
\label{split}

Figures \ref{fig:z31}(a) and (b) show  the function $M$ computed with $\tau$=5 day at 850K ($\sim 31.3$ km) just before and after the vortex split at that 
level, respectively.  Figures \ref{fig:z31}(c) and (d) display the corresponding plots for $\nabla M$, which help in the visualization of strong contrasts in 
the Lagrangian descriptor as those expected along the manifolds. Note that according to Part I, the structure of unstable and stable manifolds intersecting at 
the hyperbolic trajectory and indicated in panels (c) and (d) of these figures by the red and blue arrows, respectively, anticipates the vortex splitting at 
this level.

To visualize the process of filamentation and organization of the flow around the manifolds during the splitting, we look at the trajectories of 
particles within the kinematic vortex boundary in instances before and after the event.  In Part I, we justify our procedure to identify this 
boundary using the PDF of $M$, and argue why we expect useful information from differentiating the behaviour of particles between the regions of the kinematic 
vortex boundary that are outside and inside the curve of maximum $M$ for each longitude. Accordingly, in this case, we estimate the vortex boundary using the  
PDF of $M$ with $\tau=5$ on 9 September 2002 00:00:00 at 850K. In Figure \ref{fig:particle2Ddata}(a), the particles within the boundary are drawn in either 
blue or red according to their location inside and outside the curve defined by the maximum of values of $M$ for each longitude, respectively.  Such a 
separation helps us to capture the origin of the filaments as particles in the boundary that approach a hyperbolic trajectory after time period is equal to or 
longer than the $\tau$ used to calculate $M$.

Figure \ref{fig:particle2Ddata} shows that in  times preceding the vortex splitting, starting around 24 September 2002 09:00:00, some particles, mainly {those 
with red colour},  have been eroded from the vortex forming a long filament that extends across the south Pacific and Australia (see Figure \ref{fig:z31}(b)). 
On this day, we label selected sets of particles with capital letters in order to facilitate the description of their behaviour.

\begin{figure}
 \includegraphics[scale=0.35]{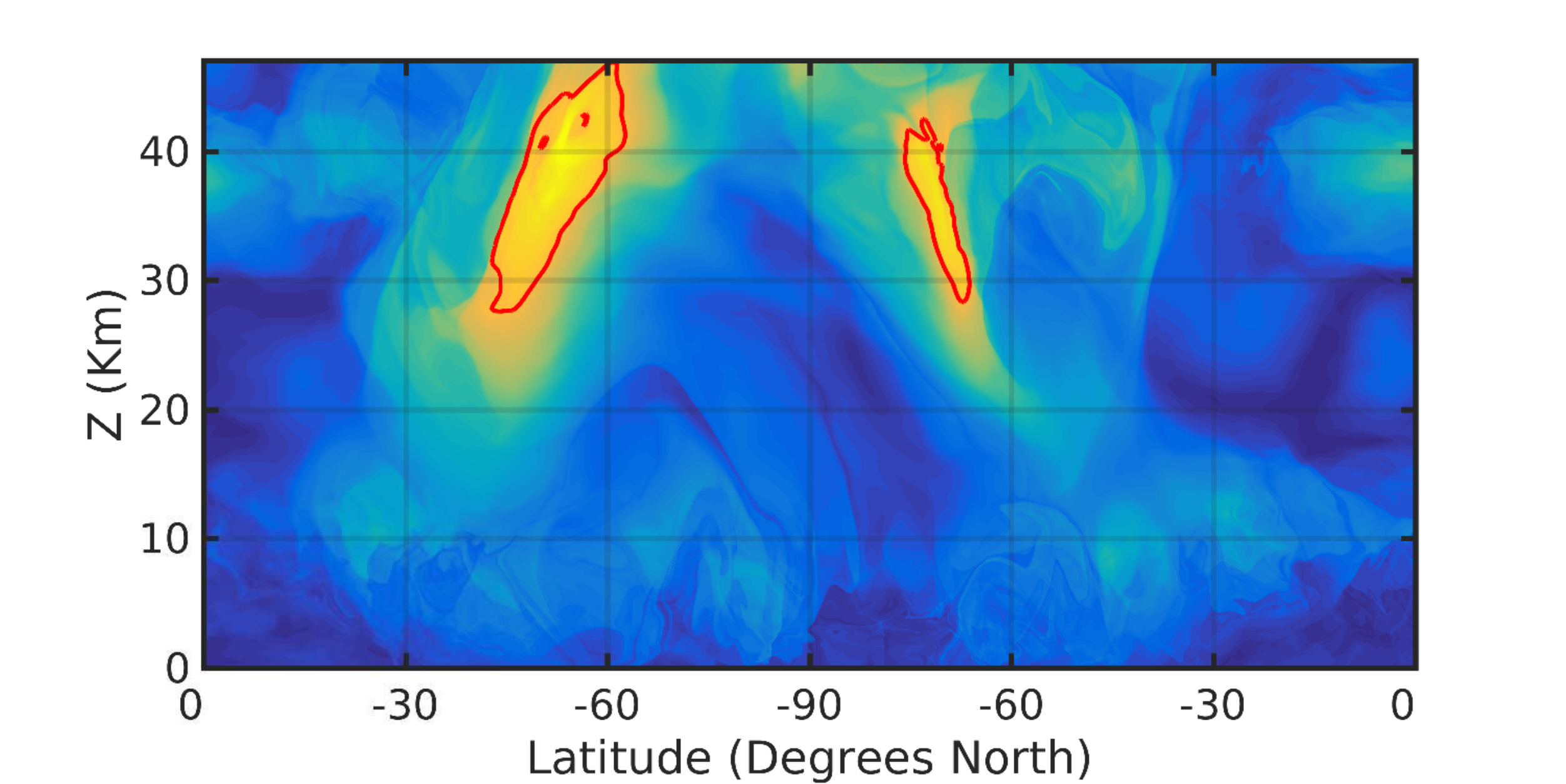}
 \centering
 \hspace{2cm} \includegraphics[scale=0.37]{colormap_M_padula_big_2.pdf}
 \caption{Vertical cross section of the function M($\tau=5$) on 9 September 2002 at $90^\circ$W and $90^\circ$E. The red lines bound the values larger than 
normalized $M=$ 0.92, which is obtained from the PDF at z=31.3km.}
 \label{fig:vertical_vortex_edge}
\end{figure}

We next inspect the role played by those manifolds highlighted by coloured arrows in Figure \ref{fig:z31} on the particle evolution during the vortex 
splitting. At this time, particles (A) over the jet approach the hyperbolic trajectory through the stable manifold (see Figs. \ref{fig:z31}(a),(c)).  Particles 
(B) and (C) are over the vortex over the South Atlantic and Indian Ocean, and both sets also approach the hyperbolic trajectory through the stable manifold  
(see Figs. \ref{fig:z31}(a),(c)).  
At later times, particles (A) move away from the hyperbolic trajectory following the branch of the unstable manifold that keeps them circling into the vortex 
over the 
South Pacific  (see Figs.\ref{fig:particle2Ddata}(b)-(d)). Particles (B) and (C) also move away from the hyperbolic trajectory following different branches of 
the unstable manifold. In this way, the set of particles (B) remains in the same vortex and the set of particles (C) swaps to the other one (see Figs. 
\ref{fig:particle2Ddata}(b)-(d)). The parcel behavior just described confirms that the thin singular lines visible in Figure \ref{fig:z31} act as 
material  barriers to transport. Additional  information on the particle evolutions in 2D is found in movie S2 of the supplementary material.

\begin{figure}[!h]
\begin{tabular}{ll}
(a) & (b)\\
\includegraphics[scale=0.7]{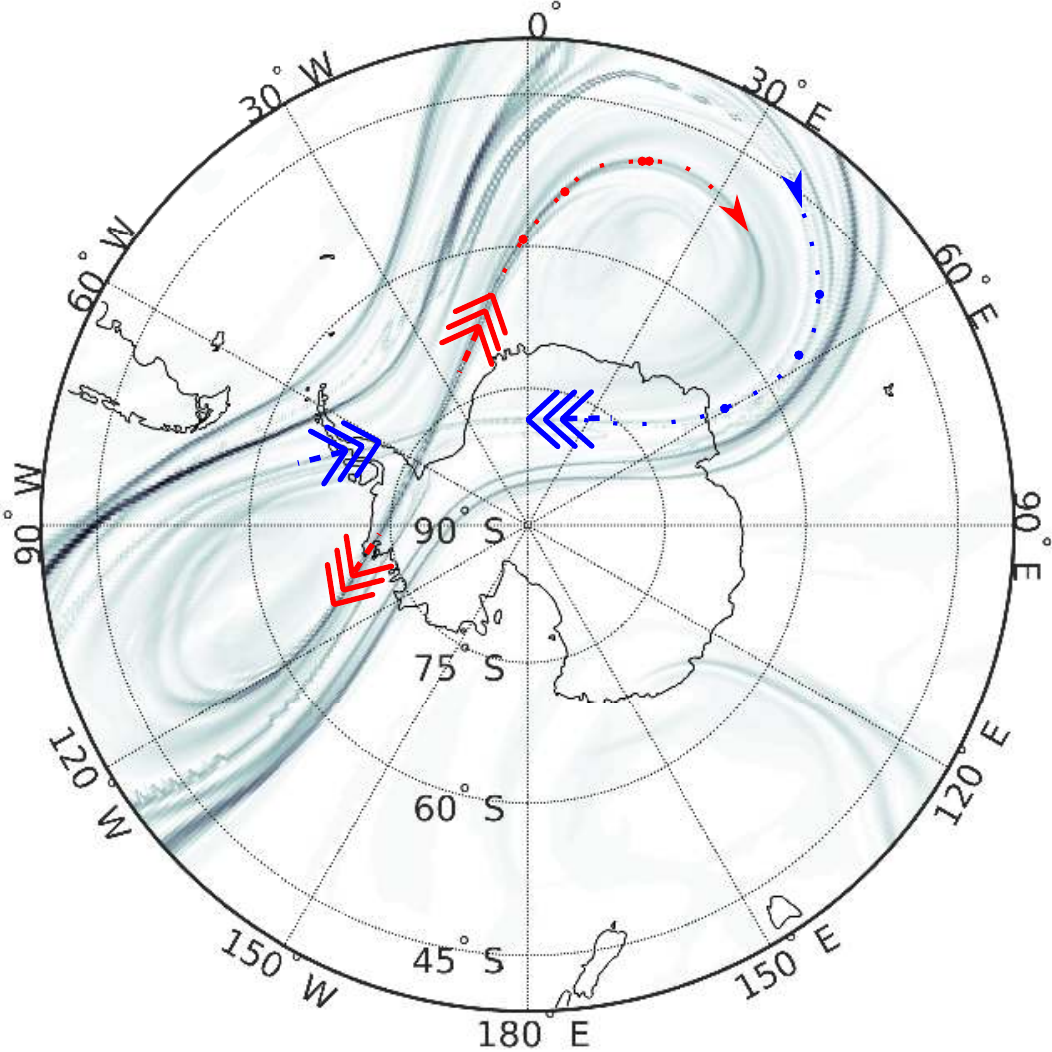}& 
 \includegraphics[scale=0.7]{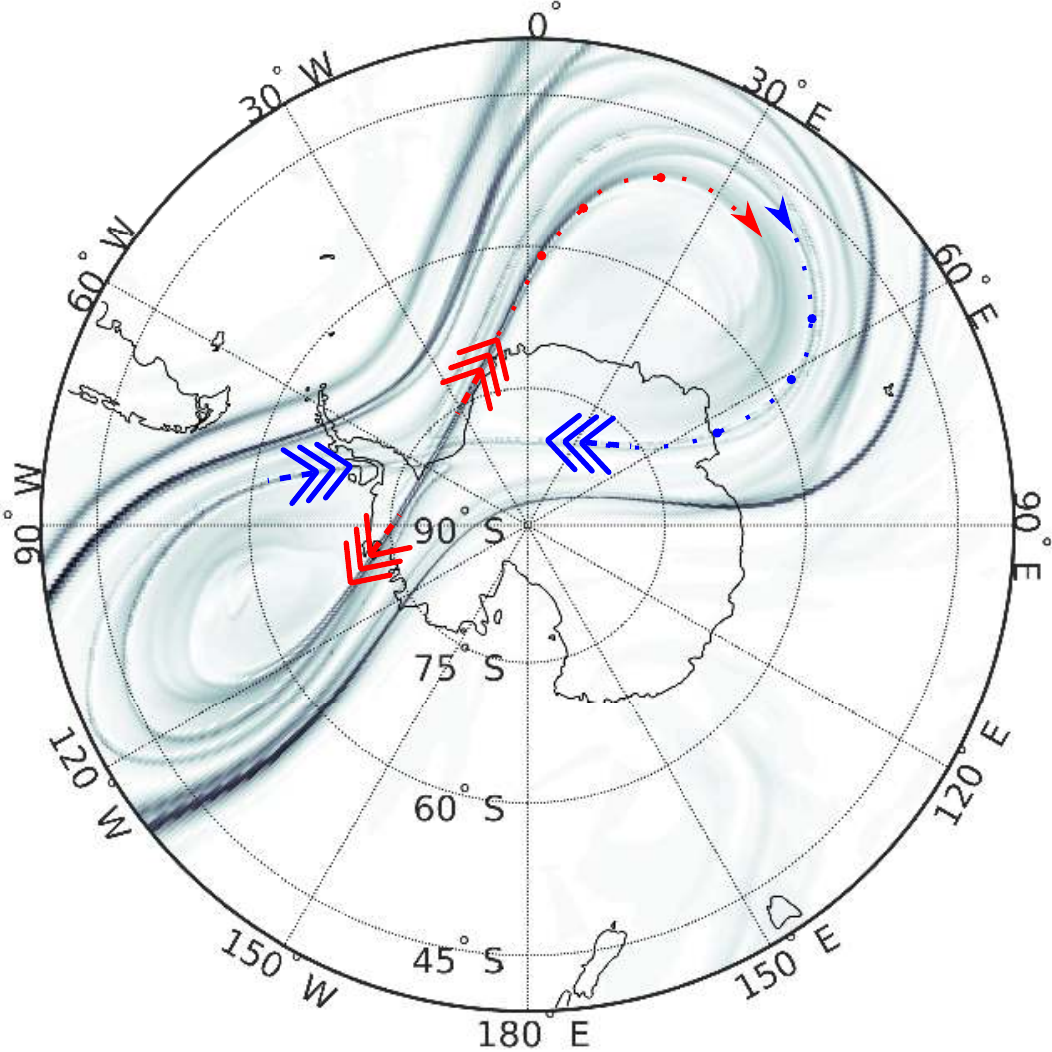}\\ 
(c) & (d)\\ 
\includegraphics[scale=0.7]{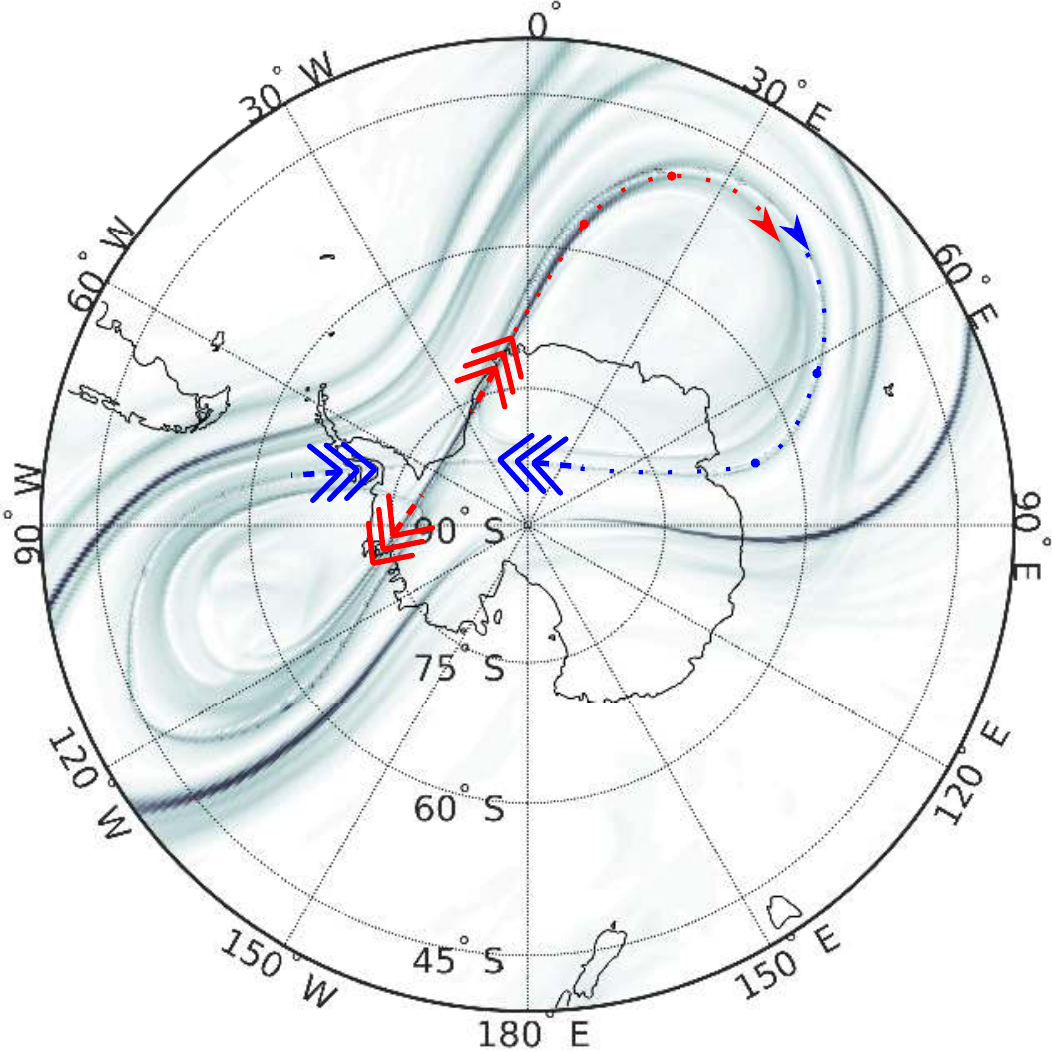}&
 \includegraphics[scale=0.7]{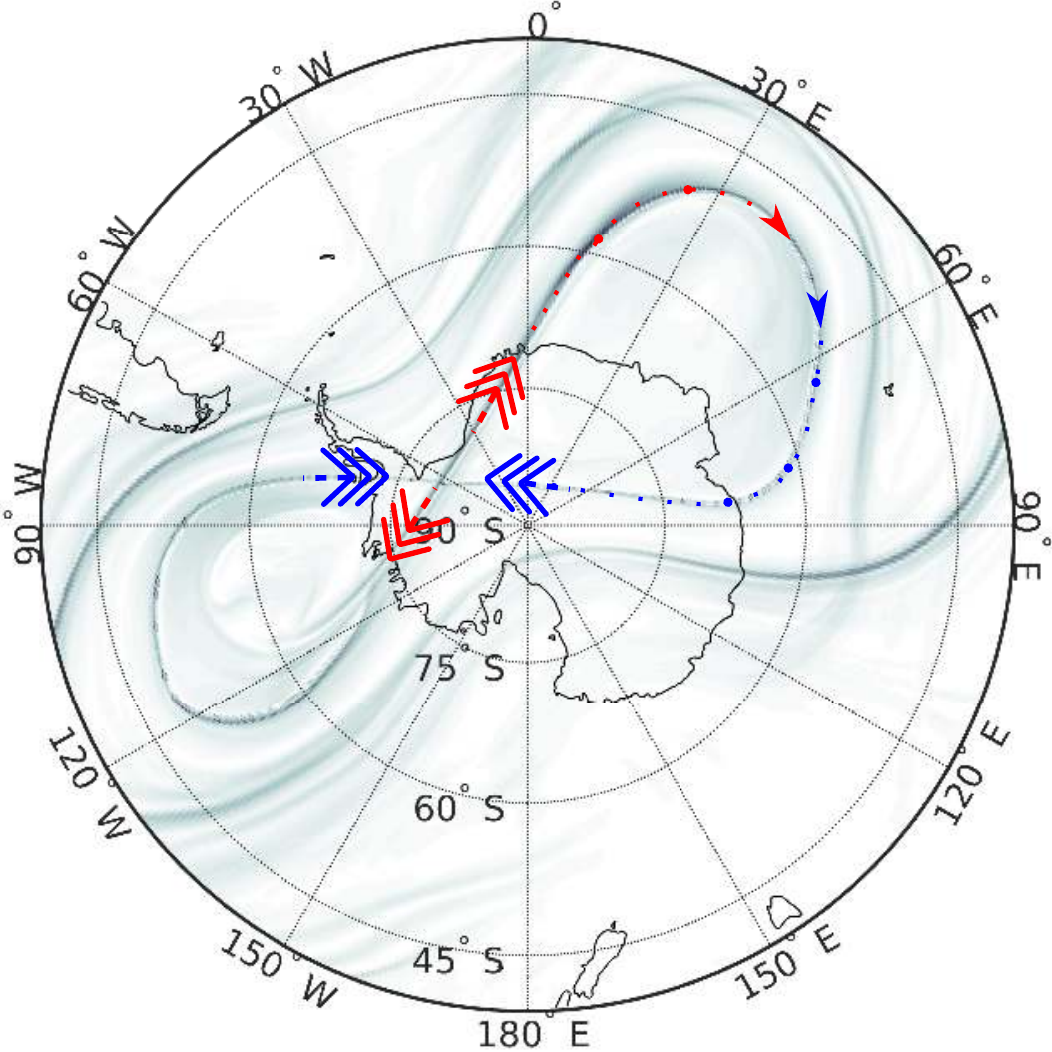}\\
\end{tabular}
\caption{$\nabla M$ ($\tau=5$) for  24 September 2002 00:00:00 at (a) 700K, (b) 600K, (c) 530K, (d) 475K}
\label{fig:grad_M_Klevel}
\end{figure}

\section{Vertical structure of the flow in the stratosphere}
\label{Vertical Structure}

To examine the vertical structure of the flow during the SPV splitting, we first extend the definition of vortex boundary from an isentropic surface as in Part 
I to a range of heights. For this, we select a {representative height} and compute the PDF of $M$ in order to obtain the value that delimits the fat tail of 
the 
distribution.  We will use the same value of $M$ for the heights within the range under consideration. For example, from {inspection of} the PDF of $M$ 
with $\tau=5$ for 9 
September 2002 00:00:00 at $z=$31.3km height, we find the normalized value $M=0.92$, which is then used for other heights. Figure 
\ref{fig:vertical_vortex_edge} shows the kinematic vortex boundary in a vertical cross-section of the function $M$ where the red lines bound the locations with 
normalized values larger than 0.92. The vortex boundary is 
several kilometers deep, extending from 27 to 42km. The particles inside these regions  produce no, or at least minimal, filamentation during the time interval 
$2\tau$ centered on 9 September 2002. Below 27 km the vortex is weaker and no clear boundary region is obtained from the procedure.

\begin{figure}[!h]
\begin{tabular}{ll}
(a) & (b) \\
\includegraphics[scale=0.4]{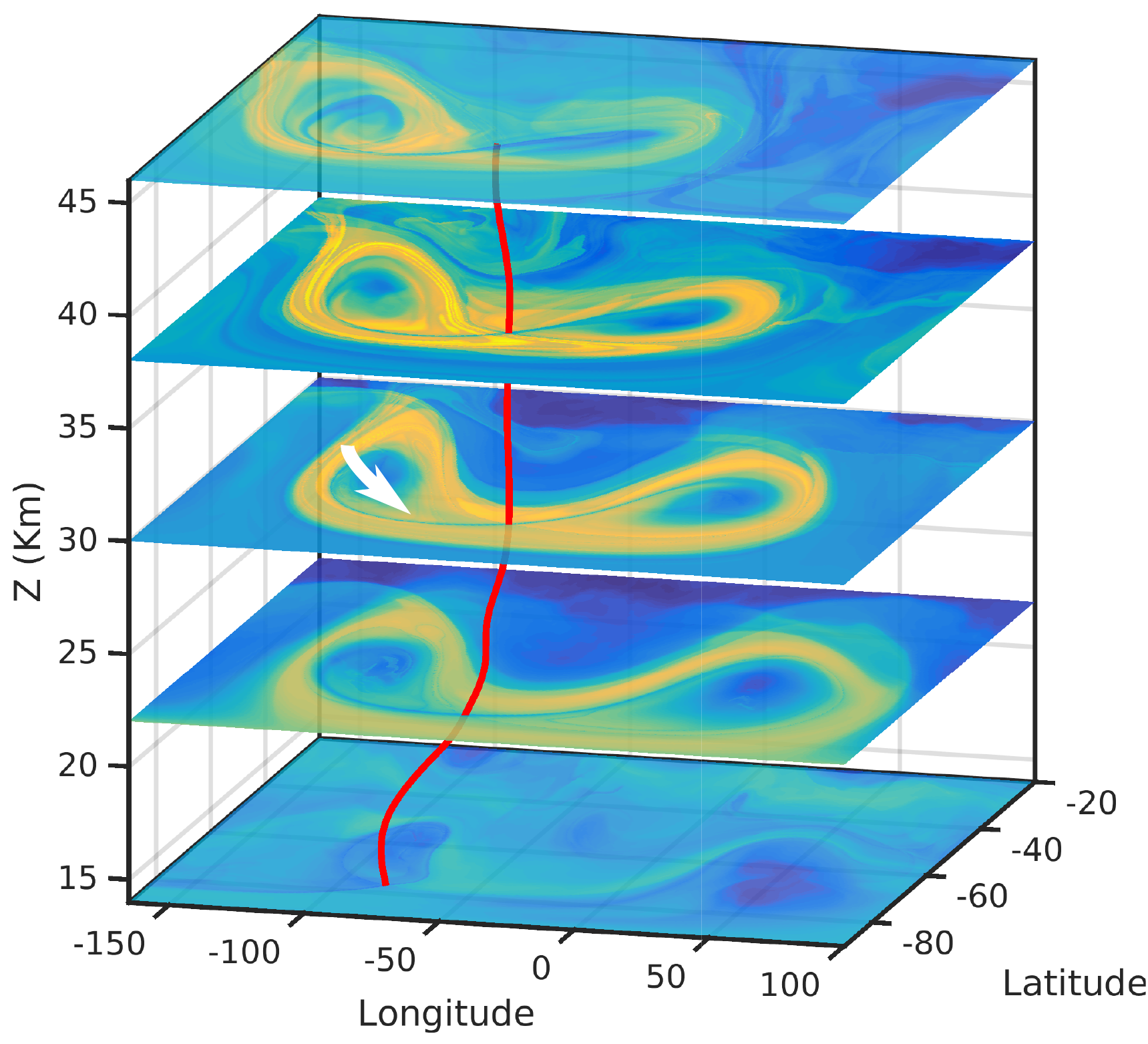} &\includegraphics[scale=0.4]{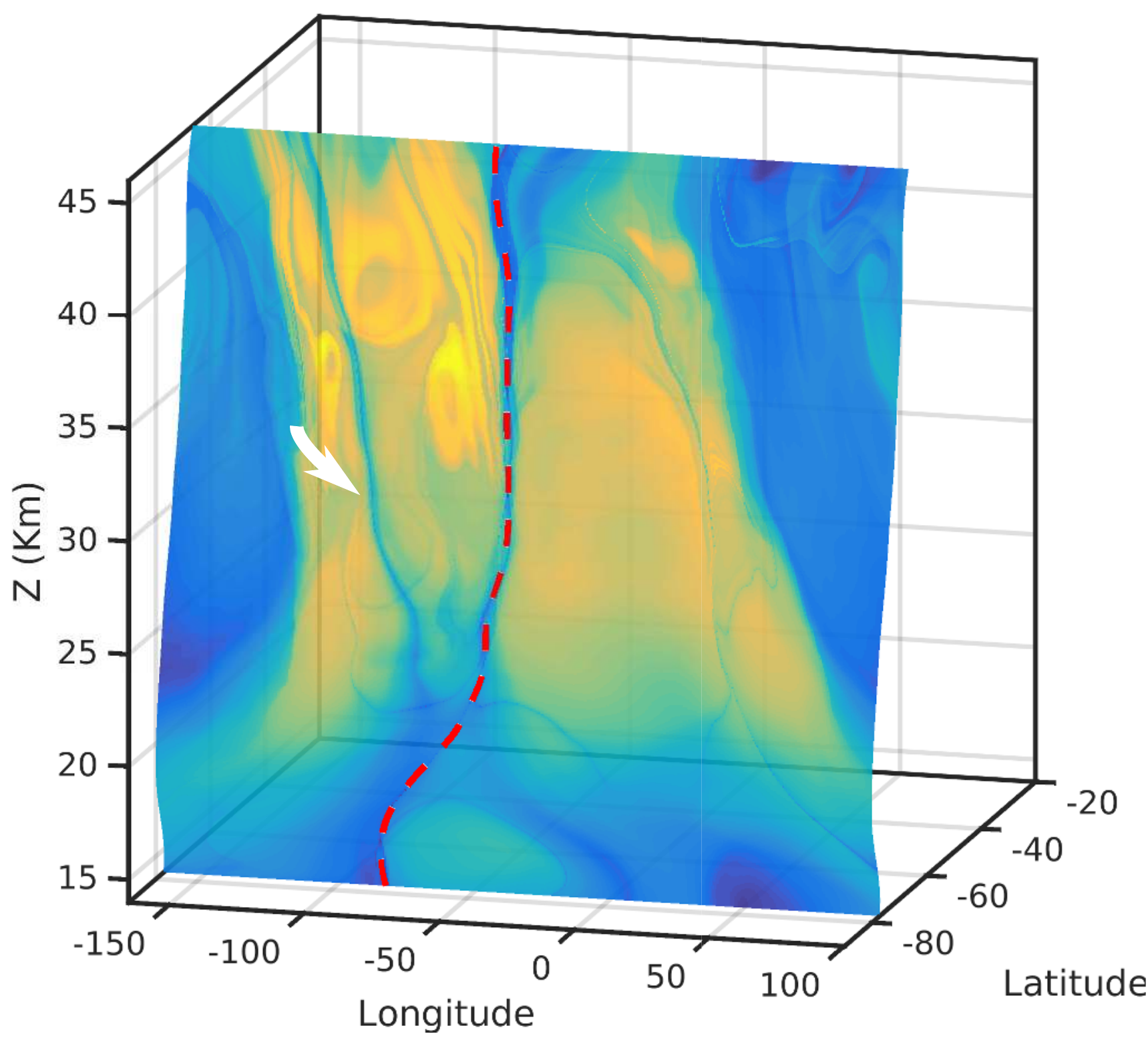}\\

\multicolumn{2}{c}{\includegraphics[scale=0.5]{colormap_M_padula.pdf}}\\
\end{tabular}
 \caption{{Structure of the polar vortex on 24 September 2002, during the SPV pinching versus longitude (degrees East) and latitude (degrees North). 
(a) Cross sections of the 
function $M$ at different horizontal levels (14, 22, 30, 38 and 46Km); the white arrow points to the unstable manifold and the NHIM is marked in red. (b) A 
surface that contains the NHIM; this surface is formed by horizontal lines of constant latitude 
 at each point of the NHIM.  Note in panel (b) that the unstable 
manifold is several km deep.} }
 \label{fig3DNHC}
\end{figure}

On 24 September 2002, during the SPV pinching displayed in Figure \ref{fig:z31}(a) at 850K, the 
unstable and stable manifolds intersect at the hyperbolic trajectory. As explained in Section 2,  
this hyperbolic 
trajectory exists at different levels conforming  the normally hyperbolic invariant curve. To show the structure of the manifolds associated with the NHIM to 
which this hyperbolic trajectory belongs, we look at cross sections of $M$ at different heights.  Figure \ref{fig:grad_M_Klevel} shows an analogue to Figure 
\ref{fig:z31}(c) at different potential temperature levels. It is clear from the figure that the hyperbolic trajectory persist  throughout all those. 

{A 3D pictorial representation of the NHIM and associated manifolds is challenging because they consist of a curve and surfaces in the 
3D space, respectively. Figure \ref{fig3DNHC} is an attempt at such a representation.   Panel \ref{fig3DNHC}(a) plots in perspective
horizontal cross-sections of $M$ at $z= 14, 22, 30, 38, 46$km, on which the stable and unstable manifolds are clearly seen. The intersections between these 
manifolds at each level define the NHIM.   This Lagrangian structure is indicated by the red line in all panels. Next we define a 3D surface  formed by 
horizontal lines of constant latitude at each point of the NHIM (see figure \ref{fig3DNHC}(b))}

{The bright yellow colors of $M$ in panels \ref{fig3DNHC}(a) and \ref{fig3DNHC}(b)  capture the SPV. Accordingly, the NHIM extends throughout a deep layer 
of the stratosphere (18 km - 45 km). The white arrows emphasize the unstable manifold in the different panels. Note in panel  \ref{fig3DNHC}(b) that the 
barrier to transport formed by the unstable manifold extends several kilometers with the vertical. }

Figure \ref{fig3DNHC} indicates that the hyperbolic trajectory can be found in a region of the stratosphere that is several km deep, i.e. the pinching event 
occurs in a wide range of heights, however the splitting does not develop at all levels but only at those in which the criterion developed in Part I is 
satisfied.
Let us now concentrate on the manifold structures below 850 K on 24 September 2002.  Figure \ref{fig:grad_M_Klevel} 
presents $\nabla M$ at 700K, 600K, 530K, and 470K on 24 September 2002 00:00:00 with $\tau=5$ days. According to Part I, the structures of unstable and stable 
manifolds intersecting at the hyperbolic trajectory and marked by 
the red and blue arrows are consistent with an evolution leading to vortex splitting only above $\sim 600 K$, which agrees with the observation.  The 
consistency is indicated by the way in which the separation between lines that contain the small blue and red arrows in Figure \ref{fig:grad_M_Klevel} changes 
in height. We can verify the applicability of our criterion by  inspecting  the trajectory of particles during the SPV splitting at different levels. The 
similarity of patterns between $\nabla M$ at different levels implies that the manifolds found are intersections of quasi-vertical surfaces. These  2D surfaces 
act as dynamical barriers in the 3D flow  and prevent transport between the emerging vortices, as confirmed in Section 4 using particles trajectories analysis.

{Although the NHIM (and hence the pinching) extends  between $\sim$ 18 km - 45 km (Figure \ref{fig3DNHC}(b)),
the vortex splitting only occurs 
from above 600K ($\sim$ 25 Km) as it can be seen by combining the information given by Figures \ref{fig:several_z} and \ref{fig:grad_M_Klevel}. This is 
consistent with the configuration of stable and unstable manifolds at each level in Figure \ref{fig3DNHC}(a) according to the splitting 
criteria in Part I, Figure 10.
We have also verified by computing parcel trajectories from different initial conditions that the unstable manifold shown in Figure \ref{fig3DNHC} acts as a 
2D barrier to the flow.
}

\section{Summary and Conclusions}
\label{s:summary}

We examine in this two-part paper the behaviour of the stratospheric polar vortex (SPV) in the southern stratosphere during the final warming in the 
{spring} of 2002. Our analysis is performed in the context of dynamical systems theory and the search for Lagrangian coherent structures: (i) hyperbolic 
trajectories and their stable and unstable manifolds, (ii) 2-tori, and (iii) the normally hyperbolic invariant manifolds (NHIM) recently identified in the 
stratospheric context. Part I presents our methodology and focuses on the understanding of fundamental processes for filamentation and ultimately for vortex 
splitting on an isentropic surface in the middle stratosphere. Part II discusses the 3D evolution of the event. In this discussion, we apply concepts developed 
in Part I concerning a definition of the vortex boundary that helps in the selection of trajectories to illuminate the  evolving flow structures, and a 
criterion that allows to anticipate at an isentropic level whereas a pinched vortex would split as it evolves in time.

\begin{figure}
\centering
 \includegraphics[scale=0.49]{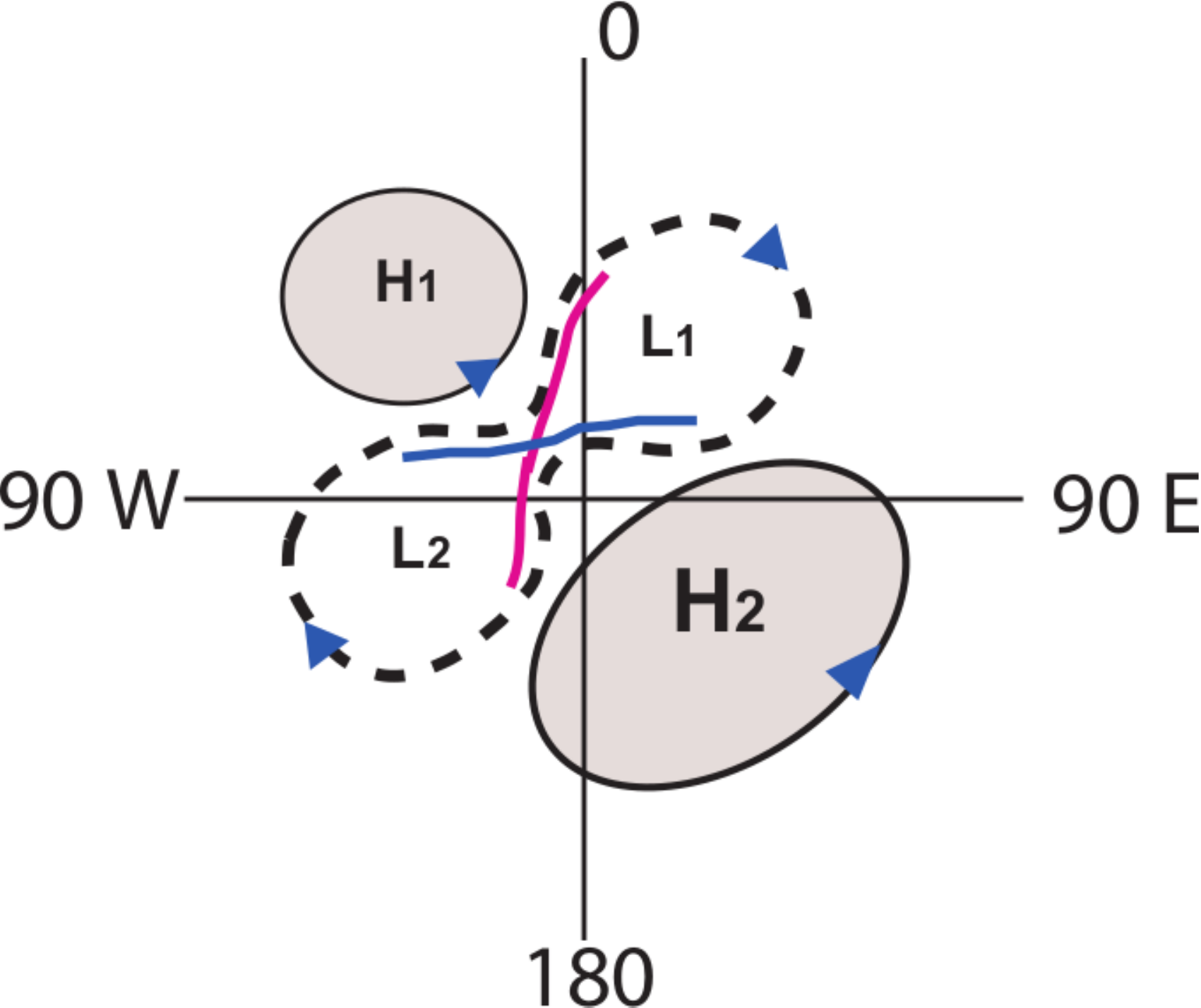}
 \caption{Schematic of the major features in the upper stratosphere during the vortex splitting in September 2002.} \label{fig:scheme}
\end{figure}

From the Lagrangian viewpoint, we have emphasized the evolution of unstable and stable manifolds, which were  crucial to the vortex splitting and essential 
components of a NHIM. Based on illustrations of the function $M$ in different cross sections we argued how the hyperbolic trajectories and intersecting 
manifolds could be thought as forming a curve and surfaces in the 3D space.  A trajectory analysis confirms that such surfaces represent barriers to the flow 
at 
least during the time corresponding to the {$\tau$} selected for calculation of $M$. We also {confirmed the consistency of the criterion based on the structure 
of manifolds on an isentropic surface developed in Part I for vortex splitting at later times and its absence below $\sim 600 K$.} 

{The 3D structure of the function $M$ shows  vortex features in the upper stratosphere during the splitting in September 2002 that
could be traced down to the troposphere.}  Figure \ref{fig:scheme} is a schematic of {such} features in the upper stratosphere. At the start of the second 
half of the September, 
the SPV was displaced from a polar position by the intensification of a quasi-stationary anticyclone south of Australia (H2), which is an element of the 
typical 
evolution of southern final warmings that generally occurs one month later in the season \cite{MOPF88,QuintanarMechoso95,Manney91,char}. The displaced SPV 
was pinched as another deep anticyclone developed  over the southern Atlantic (H1) in association with a blocking system in the troposphere, and further 
elongated (L1) as the cyclonic component of the blocking extended and tilted vertically from about $60^\circ E$ in the troposphere to the south of Africa at 
850K. Another lobe of the SPV (L2) became more sharply defined in the southeastern Pacific at 850K as Rossby wave breaking developed in the upper troposphere 
west of South America.  Starting around September 24, the elongated SPV split above approximately 600K. The lobe over the southeastern Pacific (L2) intensified 
with height, while that to the south of Africa (L1) weakened with height. After these events in late September 2002, the lobe over the western Pacific 
dissipated while the other lobe weakened and eventually became an equivalent barotropic cyclonic circulation above the South Pole in October. 

{These considerations suggest that the papers by \cite{NishiiNakamura2004} and \cite{ONeill2016} on the mechanisms for generation of the SPV 
splitting that occurred during the final southern warming of 2002 could complement each other in the following way.} \cite{NishiiNakamura2004} argued that the 
double-lobe structure of the SPV extending upwards at high latitudes in mid September 2002 (see Figure  \ref{fig:several_z}) resulted from the effects of a 
blocking system that developed in the troposphere over the southern Atlantic possibly in association with energy propagating horizontally from a burst of 
convection in the tropics. \cite{char} and \cite{ONeill2016} argued that pinching lead to vortex splitting due to vertical propagation of a disturbance that 
generated under one of the lobes of the pinched SPV.  This disturbance can be the cyclonic circulation that formed over the southeastern Pacific in association 
with Rossby wave breaking in the troposphere over the southeastern Pacific.

 \section*{Acknowledgements}

J. Curbelo and A. M. Mancho are supported by MINECO grant MTM2014-56392-R. A. M. Mancho is supported by ONR. grant No. N00014-17-1-3003. C. R. Mechoso was 
supported by the U.S. NSF grant AGS-1245069. The research of S. Wiggins is supported by ONR grant No. N00014-01-1-0769.

 \bibliographystyle{unsrt}
 \bibliography{LD}
 \vspace{1cm}
 \begin{tabular}{l}
\textbf{Jezabel Curbelo} \\
  {\small Departamento de Matem\'aticas} \\
 {\small Universidad Aut\'onoma de Madrid} \\
 {\small Instituto de Ciencias Matem\'aticas}\\
 {\small Campus de Cantoblanco UAM, 28049 Madrid, Spain.}\\
  {\small Email: jezabel.curbelo@uam.es} \\
\\

 \textbf{Carlos R. Mechoso}\\
  {\small            Department of Atmospheric and Oceanic Sciences,} \\
  {\small University of California, Los Angeles, USA.}\\
  \\
\textbf{Ana M. Mancho}\\ 
  {\small Instituto de Ciencias Matem\'aticas, CSIC-UAM-UC3M-UCM.}\\
  {\small Campus de Cantoblanco UAM, 28049 Madrid, Spain.}\\
    \\           
\textbf{Stephen Wiggins}\\
{\small School of Mathematics, University of Bristol.}\\
{\small Bristol BS8 1TW, UK.}

\end{tabular}
\end{document}